\documentclass[a4paper,11pt]{article}
\usepackage[utf8]{inputenc}
\pdfoutput=1
\usepackage{jheppub,float}
\usepackage[normalem]{ulem}
\usepackage{xcolor}
\usepackage{amsmath}
\usepackage{framed}
\colorlet{shadecolor}{lightgray}
\usepackage{hyperref}
\usepackage{parskip}

\newcommand{\p}{\partial}

\newcommand{\n}{\nabla}

\title{\boldmath Classical Soft Theorem in the AdS-Schwarzschild spacetime in small cosmological constant limit}

\author[a]{Nabamita Banerjee,}
\author[b]{Arindam Bhattacharjee,}
\author[a]{and Arpita Mitra.}

\affiliation[a]{Indian Institute of Science Education \& Research Bhopal,\\
	Bhopal Bypass Road, Bhauri, Bhopal 420 066\\ Madhya Pradesh, India.}
\affiliation[b]{Indian Institute of Science Education \& Research Pune,\\
	Homi Bhabha Road, Pashan, Pune 411 008,\\ Maharashtra, India.}\affiliation[]{E-mails:nabamita@iiserb.ac.in,
	arindam.bhattacharjee@students.iiserpune.ac.in, \\ arpitam@iiserb.ac.in}
\abstract{We have studied scattering of a probe particle by a four dimensional AdS-Schwarzschild black hole at large impact factor. Our analysis is consistent perturbatively to leading order in the AdS radius and black hole mass parameter. Next we define a proper ``soft limit" of the radiation and extract out the ``soft factor" from it. We find the correction to the well known flat space Classical Soft graviton theorem due to the presence of an AdS background.}

\begin{document} 
\maketitle
\flushbottom

\section{Introduction}
Recent developments have uncovered the central role played by soft theorems and asymptotic symmetries \cite{Campiglia:2014yka, Bern:2014vva, Broedel:2014fsa, Conde:2016csj, Laddha:2017vfh, Hamada:2017atr, Sen:2017xjn, Ashtekar:2018lor, AtulBhatkar:2018kfi, Jain:2018fda} in analyzing the vacuua of gauge and gravity theories, while probing the Infra-red structure of scattering amplitudes \cite{GellMann:1954kc, Low:1954kd, Strominger:2013jfa, Cachazo:2014fwa, Schwab:2014xua, Casali:2014xpa, Campiglia:2015yka, Chakrabarti:2017ltl, Chakrabarti:2017zmh, Zlotnikov:2017ahq, Ciafaloni:2018uwe, Addazi:2019mjh} in these theories. In the seminal work \cite{Bondi:1962px,Sachs:1962zza}, it was demonstrated that for asymptotically flat spacetimes if one recedes from sources in null directions, the symmetry algebra is not just the Poincar\'e. Rather, the asymptotic symmetry algebra, known as BMS algebra, contains the finite-dimensional Poincar\'e group as its subgroup and has an additional infinite number of generators known as supertranslations (angle dependent translations). Another important result in \cite{Bondi:1962px} was to establish that the mass loss due to gravitational radiation is a nonlinear effect of general relativity and the emission of gravitational waves from an isolated system is accompanied by a mass loss from the system.

In general relativity the classical vacuum is highly degenerate. Different vacua, which spontaneously break BMS symmetry, are related to each other by `supertranslations'. In the quantum picture, these vacua differ by addition of a soft gravitons. In the soft (zero frequency) limit the scattering amplitude involving hard particles and soft particles can be recast as a (soft) factor multiplied with the scattering amplitude of only hard particles. This is the celebrated Soft theorem by Weinberg \cite{Weinberg:1964ew, Weinberg:1965nx}, which relates an amplitude with soft photons or/and gravitons to amplitudes without any soft particle, in a quantum theory of photons and gravitons in asymptotically flat space times. Soft (photon) graviton theorem can be recast as the Ward identities for (large $U(1)$ gauge transformations for QED \cite{He:2014cra, Lysov:2014csa, Kapec:2014zla, Campiglia:2019wxe}) BMS transformations for gravity \cite{He:2014laa, Avery:2016zce, Hamada:2018vrw, AtulBhatkar:2019vcb} in the asymptotic limit, i.e. when we take the radial distance to infinity or equivalently consider early and late retarded time. The soft factor is universal in the leading order in  frequency (of soft particles) expansion. This implies that the internal structure of the objects involved in a scattering as well as the details of their interactions are irrelevant in the large wavelength or equivalently zero frequency limit. The zero frequency limit was initially considered in \cite{Smarr:1977fy}, where  the energy spectrum ($dE/d\omega$) was shown to be independent of frequency $\omega$ if the asymptotic trajectories have constant velocity, at least one of them being nonzero. This implies an enhancement of symmetries in the asymptotic limit at low frequencies.

An interesting development in this field of research has come in last two years with works by Sen et.al \cite{Laddha:2018rle, Laddha:2018myi, Laddha:2019yaj, Saha:2019tub}. First in \cite{Laddha:2018myi} they showed that in larger than four spacetime dimensions, the soft factors determine
the low frequency radiative part of the fields that are produced during a classical scattering in electromagnetic and gauge theories. Moreover in four dimensions, although the S-matrix is IR divergent, the radiative part of the fields provides an unambiguous definition of soft factor in an appropriate classical limit. Thus they framed an alternate way to find the soft factor for gauge and gravity theories in asymptotically flat spacetimes. The soft expansion of the radiative field defines ``Classical Soft" theorems. Sen et. al. independently proved these theorems by a classical analysis in \cite{Laddha:2018rle} and \cite{Saha:2019tub}. Both the photon and graviton soft factors for asymptotically flat spacetime can be used to determine the memory effect and a tail term to it arising from scattering processes involving several outgoing light particles and no incoming light particles \cite{Laddha:2018vbn, Sahoo:2018lxl, Fernandes:2020tsq}. Although their works are primarily valid for classical scattering computations in theories in asymptotically flat spacetimes, but the idea can be extended to other curved spaces as well, that does not behave as flat at large distances. This is the key idea of this present work, where we address the issue of soft factor in theories in asymptotically AdS spaces by studying classical radiation (bremsstrahlung) \cite{DeWitt:1960fc, Peters:1966, Peters:1970mx, Kovacs:1977uw} in a AdS Schwarzschild background.   

Particle trajectories in AdS spacetime behave like particle in a box as the null rays get bounced back from time like boundary infinite number of times. Thus, for a quantum field theory defined in an asymptotically AdS space, the usual notion of ``in" and ``out" states does not exist. This is the prime obstruction in defining the usual scattering amplitudes in such theories\footnote{for scaterring in AdS we refer to \cite{Gary:2009mi, Penedones:2010ue, Fitzpatrick:2011ia, Rastelli:2016nze}}. As a result, unlike asymptotically flat spacetimes, a soft theorem or its analogue is not presently known for quantum field theories defined in an asymptotically AdS spacetime\footnote{a related work can be found in \cite{Hijano:2020szl}}. Therefore using the formulation of classical soft theorem we can identify the analogue of soft factors for classical scattering in AdS backgrounds. For technical simplification, we consider the value of cosmological constant to be small and treat it as a perturbation parameter for our computations. Thus our result provide us corrections to the known results of Classical Soft Theorem in asymptotically flat Schwarzschild case, to the linear order in cosmological constant. For our purpose, we consider scattering of a probe of mass $m$ by a scatterer of mass $M$ (the black hole)
due to gravitational interaction in the asymptotically AdS space, where $M>>m$. 
We further assume that the distance of closest approach between the
probe and the scatterer, i.e. the impact parameter, is large compared to the Schwarzschild radius of the scatterer. All our results are valid to the
linear order in the ratio of the Schwarzschild radius and the impact parameter.
Finally we study the ``soft limit" of the gravitational radiation coming out of the above scattering process. Since we are working in asymptotically AdS spacetime, we need to be careful about the consideration of soft limit. For gravitational radiation in asymptotically flat theories, soft limit implies vanishing of the four momentum of the radiation, and hence vanishing of its frequency. But a radiation mode inside the AdS spacetime have the minimum frequency inversely proportional to the size of the spacetime. Thus consideration of a limit of strictly vanishing frequency (of radiations) is not possible in asymptotically AdS spaces. However we may consider the limit where frequency of radiation and cosmological constant simultaneously tend to zero, keeping there ratio finite. Physically it implies that we consider the radiation to limit to a strictly soft one as the space is limiting to a flat space.  We show that the radiation can leak from the spacelike infinity because of the polynomial potential that arises due to the small cosmological constant.
 
The paper is organized as follows. In section \ref{sc2} and section \ref{sc3}  we study properties of four dimensional AdS Schwarzschild metric and write it in the isotropic coordinates that is useful for further computations of radiation. Our computations are valid upto linear order in black hole mass parameter $M$ and square of inverse AdS radius $\frac{1}{l^2}$.  In section \ref{sc4} we set up the scattering problem due to a probe mass $m$ in the AdS Schwarzschild background. To find the radiation profile, we need to solve Einstein's equations. In section \ref{sc5} we set up the technical details to achieve the same. In section \ref{sc6}, we provide the solution to classical radiation and express the same as function of its frequency and position by performing a Fourier transform in the time coordinate. We then carry out the Laurent series expansion in the frequency of soft radiation for the classical radiative field in the Fourier space in section \ref{sc7}. This section contains the prime result of the present paper. Here we derive the modification to the Classical soft theorem due to the AdS potential. We finally conclude the paper in section \ref{sc8}. The two appendices contain some computational details. 

\section{AdS-Schwarzschild metric}\label{sc2}
In this section we study perturbations of a four dimensional theory of gravity with negative cosmological constant $\Lambda$, induced by a point mass moving in an unbound trajectory. The gravitational action for this case is given by
\begin{equation}
S=\int d^4 x \sqrt{-g}(R-2\Lambda).
\end{equation}
The equations of motion for the metric tensor is,
\begin{equation}
R_{\mu\nu}-\frac{1}{2}R g_{\mu\nu}+\Lambda g_{\mu\nu}=0\label{em}.
\end{equation}
Considering the solutions of these equations (\ref{em}) for a static spherically symmetric spacetime with mass M and a negative cosmological constant $\Lambda=-\frac{3}{l^2}$ we can write the metric as \cite{Cruz:2004ts}, 
\begin{equation}
ds^2=-f(r)dt^2+\frac{d r^2}{f(r)}+r^2(d\theta^2+\sin^2 \theta d\phi^2)\label{metric},
\end{equation}
where the lapse function $f(r)$ in global coordinates is
\begin{equation}
f(r)=1-\frac{2GM}{r}-\Lambda\frac{r^2}{3}=1-\frac{2GM}{r}+\frac{r^2}{l^2}
\end{equation}
The range of the coordinates involved in the metric are 
\begin{equation}
-\infty \leq t\leq \infty; r\geq 0; 0\leq \theta\leq \pi; 0\leq\phi\leq 2\pi.
\end{equation}
We consider mostly plus convention for the metric. The lapse function vanishes at the horizon values of `$r$' which are the roots of $r^3+l^2 r-2Ml^2=0$ (in $G=1$ unit).  For negative cosmological constant, the metric function has only one real positive root \cite{Cruz:2004ts}. Thus the radius of horizon is, 
\begin{equation}
r_h=\frac{2}{\sqrt{3}}l\sinh\left[\frac{1}{3}\sinh^{-1}(3\sqrt{3}\frac{M}{l})\right].
\end{equation}   
Expanding this in terms of `$M$' with $1/l^2\ll M^2/9$ we obtain
\begin{equation}
r_h \approx 2M-\frac{8M^3}{l^2}+...
\end{equation}
\\
Asymptotically the metric (\ref{metric}) reduces to the general AdS metric
\begin{equation}
ds^2=-\left(1+\frac{r^2}{l^2}\right)dt^2+\frac{dr^2}{\left(1+\frac{r^2}{l^2}\right)}+r^2(d\theta^2+\sin^2 \theta d\phi^2).
\end{equation}

\begin{figure}
\hspace*{90pt}
\includegraphics[scale=0.5]{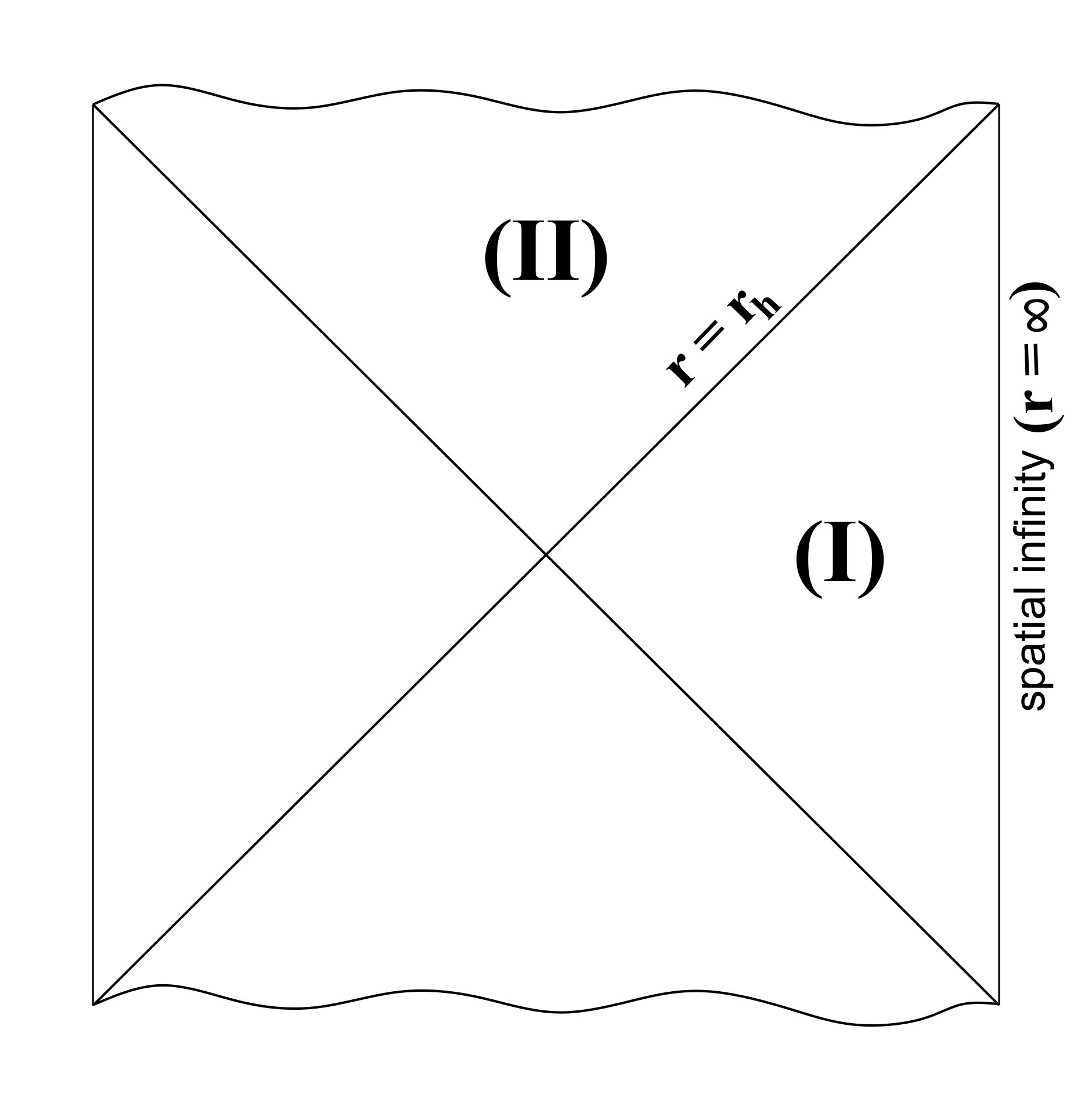}
\caption{Penrose diagram for AdS-Schwarzschild Metric. The structure at spatial infinity is different here than Schwarzschild black hole in flat space. The wiggly lines represent the black hole singularity at $r = 0$. The probe scattering occurs at region (I).}
\label{penrose}
\end{figure}

As $l^2\geq 0$, $1+\frac{r^2}{l^2}$ has no real roots. Therefore this metric supports no cosmological horizon. A massless particle, e.g. a photon, can reach the spacelike boundary of the above spacetime in finite time. While massive particles, moving along geodesics, can never get to the boundary. Thus it is possible to send a beam of light to infinity, and have it come back in a finite amount of time \cite{Aharony:1999ti}. If we consider a radial ($d\theta=d\phi=0$) null ray ($ds^2=0$) from the origin to spatial infinity, the time taken to reach the spatial infinity is
\begin{align}
t=\int_0^R\frac{dr}{1+\frac{r^2}{l^2}}=l \arctan(R/l)\label{timet},
\end{align}  
where $R$ denotes the position of spatial infinity. We have given the Penrose diagram for this space above in figure \ref{penrose}. From (\ref{timet}) we notice that, when $R\rightarrow \infty$; $t\rightarrow\frac{l\pi}{2}$. Thus for small $l$, the time taken is small. It is necessary to impose special boundary conditions at the spatial infinity, $r=\infty$, to make sure that the fields reach there. The common constraint to impose is an energy conservation such that the energy does not leak out of the universe. The energy conservation leads to the reflective boundary condition at spatial infinity for fields at the AdS boundary
. This condition implies a non-vanishing self-force unlike in deSitter space, where the self-force vanishes. Alternately this happens because AdS is conformal not to Minkowski space, but to Minkowski space with special boundary conditions. However in our calculation we consider large $l$ limit. For large $l$ and $R\ll l$, the time taken to reach the spatial infinity is comparable to $R$. Thus reflective behaviour of the boundary does not come into play to this order. We elaborate more on this in section \ref{sc7}.

\section{Isotropic coordinates}\label{sc3}
In this section we introduce the isotropic coordinates for the metric (\ref{metric}). Isotropic coordinate system is useful in computing perturbations. 
Since the metric (\ref{metric}) in isotropic coordinates has its spacelike slices conformal to the Euclidean one, the resulting radiative components of gravitational perturbations are also isotropic in all spatial directions. We truncate the metric components to order ${\cal{O}}(1/l^2)$ during the derivation of isotropic coordinates. Throughout this paper, we consider $r^2/l^2\ll1$ limit. The resulting metric describes spatial slices of AdS only up to this order. Hence all calculations will be carried out up to this order. During the computation of gravitational perturbation we also assume the probe scattering limit. This limit considers the point particle to have a large impact parameter from the central black hole i.e. $M/r\ll 1$.  These two assumptions together allow us to discard the terms proportional to $M^2/l^2$, $Mr/l^2$ and other higher order terms throughout the calculation. Therefore we always consider terms up to ${\cal{O}}(M)$ and ${\cal{O}}(1/l^2)$.

To proceed we rewrite (\ref{metric}) as
\begin{align}
d\tilde{s}^2=-\left(1-\frac{2GM}{r}+\frac{r^2}{l^2}\right)dt^2+B^2(\rho)(d\rho^2+\rho^2d\Omega^2).\label{mm}
\end{align}
The spatial part and angular part of (\ref{mm}) are related to the same of (\ref{metric}) as follows,
\begin{align}
B^2(\rho)d\rho^2=\frac{dr^2}{1-\frac{2GM}{r}+\frac{r^2}{l^2}};\qquad B^2(\rho)\rho^2=r^2\label{B}.
\end{align}
These two equations lead to the relation between Schwarzschild coordinates and isotropic coordinates up to ${\cal{O}}(M)$ and ${\cal{O}}(1/l^2)$ as,
\begin{align}
\frac{d\rho}{\rho}&=\ln\left(r-GM+r\sqrt{1-\frac{2GM}{r}}\right)-\frac{r^2}{4l^2}+\cdots\notag\\\Rightarrow\rho&=r\left(1-\frac{GM}{r}-\frac{r^2}{4l^2}\right).
\end{align}
The above relation implies that the Schwarzschild coordinate `$r$' is related to the isotropic coordinate '$\rho$' as,
\begin{equation}
r\sim \rho\left(1+\frac{GM}{\rho}+\frac{\rho^2}{4l^2}\right)\label{reln}.
\end{equation}
and in isotropic coordinates (\ref{metric}) becomes,
\begin{equation}
ds^2=g_{00}dt^2+g_{ij}dx^idx^j,
\end{equation}
where,
\begin{align}
g_{00}\approx -\left(1+2\phi+\frac{\rho^2}{l^2}\right), \qquad g_{ij}\approx \delta_{ij}\left(1-2\phi+\frac{\rho^2}{2l^2}\right)\label{metriciso}.
\end{align}
Here $\phi=-\frac{GM}{\rho}$ is the gravitational potential due to the black hole mass $M$ and we have considered the limit $\rho^2/l^2\ll 1$.
Similarly, the inverse metric components are given by,
\begin{align}
g^{00}\approx -\left(1-2\phi-\frac{\rho^2}{l^2}\right), \qquad g^{ij}\approx \delta^{ij}\left(1+2\phi-\frac{\rho^2}{2l^2}\right)\label{inmetriciso},
\end{align}
and the square root of the determinant of the metric is 
$ \quad\sqrt{-g}=1-2\phi+\frac{5\rho^2}{4l^2}$.
 
We have given the expressions for non-vanishing connection, Riemann tensor, Ricci tensor and Ricci scalar derived using the metric (\ref{metriciso}) in the appendix \ref{AppA}.

\section{Probe scattering in AdS Schwarzschild background}\label{sc4}

In this section, we study the effect of a probe mass on the AdS Schwarzschild background, that we have given in the last section. To achieve this, we linearly perturb the background spacetime by introducing a point particle of mass $m$ with a stress tensor $T^{\mu \nu}_{(P)}$ :
\begin{align}
T^{\mu \nu}_{(P)} =m\int \delta(x,z(s))\frac{dz^{\mu}}{ds}\frac{dz^{\nu}}{ds}\,ds\label{ps}.
\end{align}
Here $\delta(x,z(s))$ corresponds to the covariant delta function normalized as
\begin{equation}
\int \sqrt{-g}~\delta(x,z(s)) ds =1.
\end{equation}
This induces a perturbation of the metric as $
g_{\mu\nu} \to g_{\mu\nu} + h_{\mu \nu}$. Under this perturbation we take the variation of (\ref{em}) to linear order in $h_{\mu \nu}$ as
\begin{align}
\nabla^{\rho}\nabla_{\nu}\bar{h}_{\mu\rho}+\nabla^{\rho}\nabla_{\mu}\bar{h}_{\nu\rho}&-\bar{h}_{\mu\nu;\alpha}{}^{\alpha}-g_{\mu\nu}\nabla^{\rho}\nabla^{\alpha}\bar{h}_{\alpha\rho}-(R-2\Lambda)\bar{h}_{\mu\nu}-\Lambda g_{\mu\nu}\bar{h}+g_{\mu\nu}R^{\alpha\beta}\bar{h}_{\alpha\beta}\notag\\&=16\pi G\delta T_{\mu\nu}.\label{eqhbar}
\end{align}
Here $\delta T_{\mu\nu}=T_{\mu\nu}{(P)}$ and $\bar{h}_{\mu\nu}$ is the trace reversed metric perturbation defined as 
\begin{equation}
\bar{h}_{\mu\nu}=h_{\mu\nu}-\frac{1}{2}g_{\mu\nu}h\label{trmp}
\end{equation} 
with $h$ being the trace of perturbation. Simplification of  (\ref{eqhbar}) can be done by introducing a gauge fixing vector function $f$ in the following way, 
\begin{equation}
 \nabla^{\alpha}\bar{h}_{\alpha\beta}=f_{\beta}.
 \label{gauge}
\end{equation}
Using the identity
\begin{equation}
\nabla^{\rho}\nabla_{\nu}\bar{h}_{\mu\rho}=\nabla_{\nu}\nabla^{\rho}\bar{h}_{\mu\rho}+R^{\sigma}{}_{\mu\nu}{}^{\rho}\bar{h}_{\sigma\rho}+R^{\sigma}{}_{\nu}\bar{h}_{\mu\sigma},
\end{equation} 
(\ref{eqhbar}) boils down to,
\begin{align}
\bar{h}_{\mu\nu;\alpha}{}^{\alpha}&-2 R^{\sigma}{}_{\mu\nu}{}^{\rho}\bar{h}_{\sigma\rho}-2 R^{\sigma}{}_{(\mu}\bar{h}_{\nu)\sigma}-g_{\mu\nu}R^{\alpha\beta}h_{\alpha\beta}+(R+\frac{6}{l^2})\bar{h}_{\mu\nu}-\frac{3}{l^2}g_{\mu\nu}\bar{h}\notag\\&-2(\nabla_{(\mu}f_{\nu)})+g_{\mu\nu}\nabla^{\alpha}f_{\alpha}=-16\pi G \delta T_{\mu\nu}\label{eom1}.
\end{align}
We first compute the spatial components $\bar{h}_{ij}$ of the radiative gravitational perturbation. Once $\bar{h}_{ij}$ is known, by the gauge condition (\ref{gauge}), one can then easily compute $\bar{h}_{00}$ and $\bar{h}_{0i}$. Finally we choose an explicit gauge fixing condition on $f_{\beta}$ as,
\begin{equation}
f_{\beta}=-2\left(\phi_k-\frac{\rho_k^2}{4l^2}\right)\bar{h}_{k\beta}
\label{FE}.
\end{equation}
This choice simplifies the gauge terms of (\ref{eom1}) in further computations. Putting $\mu=i$, $\nu=j$ in (\ref{eom1}) we evaluate each term as,
\begin{align*}
&-(f_{i,j}+f_{j,i}+\delta_{ij}f_{0,0}-\delta_{ij}f_{k,k})\notag\\&=2(\phi_{kj}-\frac{\rho_{kj}^2}{4l^2})\bar{h}_{ki}+2(\phi_{k}-\frac{\rho_{k}^2}{4l^2})\bar{h}_{ki,j}+2(\phi_{ki}-\frac{\rho_{ki}^2}{4l^2})\bar{h}_{kj}+2(\phi_{k}-\frac{\rho_{k}^2}{4l^2})\bar{h}_{kj,i}\notag\\& +2\delta_{ij}(\phi_{k}-\frac{\rho_{k}^2}{4l^2})\bar{h}_{k0,0}-2\delta_{ij}(\phi_{l}-\frac{\rho_{l}^2}{4l^2})\bar{h}_{kl,k}-2\delta_{ij}(\phi_{lk}-\frac{\rho_{lk}^2}{4l^2})\bar{h}_{lk}.
\end{align*}
Therefore manipulating terms in (\ref{eom1}) we finally get the following equation,
\begin{align}
&\Box((1+2\phi-\frac{\rho^2}{2l^2})\bar{h}_{ij})+4\left[\phi\bar{h}_{ij,00}+\phi_{i}\bar{h}_{j0,0}+\phi_{j}\bar{h}_{i0,0}+\frac{1}{2}(\phi_{ij}-\frac{\delta_{ij}}{2}\phi_{kk})(\bar{h}_{00}+\bar{h}_{ll})\right]\notag\\&+\frac{1}{2l^2}\left[\rho^2 \bar{h}_{ij,00}+\rho^2_{i}\bar{h}_{j0,0}+\rho^2_{j}\bar{h}_{i0,0}+\rho_{ij}^2(2\bar{h}_{00}-\bar{h}_{ll})+\frac{3}{2}\delta_{ij}\rho^2_{kl}\bar{h}_{kl}-\frac{1}{2}\delta_{ij}\rho^2_{kk}\bar{h}_{ll}\right.\notag\\&\left.+\frac{3}{2}(\rho^2_{ki}\bar{h}_{kj}+\rho^2_{kj}\bar{h}_{ki})-\rho^2_{kk}\bar{h}_{ij}+\frac{3}{2}\rho_k^2\bar{h}_{ij,k}\right]=-16\pi G\delta T_{ij}\label{maineom},
\end{align}
where $\delta T_{ij}$ follows from (\ref{ps}),
\begin{equation}
\delta T_{ij}=g_{ik}g_{jm}\delta T^{km}=m ~\left(1-4\phi+\frac{\rho^2}{l^2}\right)\int \delta^4(x,z(s))\frac{dz^{i}}{ds}\frac{dz^{j}}{ds}\,ds
\end{equation}
and we define $\Box=\eta^{\mu\nu}\p_{\mu}\p_{\nu}=-\p_0^2+\p_i^2$. Here, we have expressed (\ref{maineom}) as the differential equation involving the flat spacetime D'Alembertian operator and other additional terms involving $\phi$, $\rho^2/l^2$ and their derivatives. This helps us to solve the equation perturbatively in zeroth order using the Green's function corresponding to the flat spacetime D'Alembertian operator. We eventually find the solution upto ${\cal{O}}(M)$ and ${\cal{O}}(1/l^2)$ plugging the zeroth order solution in the source terms of (\ref{maineom}). 

\section{Scalar Green's function in Anti-deSitter space}\label{sc5}

The prime technical aspect of this paper is to first find a solution of (\ref{maineom}). As it turns out, this is much easier to achieve using Synge's worldline formalism. We refer the readers to \cite{J.L.Synge:1960zz} for a detailed analysis of the same. In this formalism we express the solution for the perturbation in terms of a well known quantity known as Synge's world function, that we introduce in this section. Although our goal is to find solution of equation (\ref{maineom}), we first proceed with a simpler case, namely the scalar box equation. In its integral form, the equation takes the following form, 
\begin{equation}
\psi_{;\alpha}{}^{\alpha}(x) = -\int \delta(x,z(s)) f(s) ds\,.
\label{scalar.cov}
\end{equation}
The form of $f(s)$ depends on the sources appearing in the equation and `$_;$' denotes covariant derivative with respect to any background geometry. Comparing the above equation with (\ref{eom1}), we do not have a nontrivial tensor structure and the terms proportional to curvature tensors and gauge fixing function are absent in this case.

The covariant delta function $\delta(x,z(s))$ is related to the flat spacetime delta function $\delta^{4}\left(x-z(s)\right)$ via
\begin{equation}
\delta(x,z(s)) \sqrt{-g} = \delta^{4}\left(x-z(s)\right) = \delta(t - z^0(s)) \delta^{(3)}(\vec{x} - \vec{z}(s))\,.
\end{equation}
In flat spacetime, (\ref{scalar.cov}) becomes $\Box \psi(x)=-J(x)$. The solution of $\psi(x)$ is,
\begin{align}
\psi(x)=\int G(x,y)J(y) d^4 y &=\int G(x,y)~\delta^4(y,z(s))f(s) ds \, d^4 y\notag\\\quad &=\int G(x,z(s))f(s)ds,
\end{align}
where $J$ is the source and $f(s)$ is strength of the source. The flat Green's function $G$ satisfies,
\begin{equation}
\Box G\left(x,z\right) =\eta^{\mu\nu}\partial_{\mu}\partial_{\nu} G\left(x,z\right) = -\delta^{4}\left(x-z\right).
\end{equation}
As is well known,the flat space Green function has 3 possible solutions: 
\begin{align}
{\text {the retarded one:} } \qquad G_R(x,z) = \frac{1}{4\pi \vert R \vert} \delta\left(t - z_0 - \vert \vec{R}\vert \right),
\end{align}
\begin{align}
 {\text {the advanced one:} } \qquad G_A(x,z) = \frac{1}{4\pi \vert\vec{R} \vert} \delta\left(t - z_0 + \vert \vec{R}\vert \right)
\end{align}
and a sum over retarded and advanced Green functions
\begin{align}
G(x,z) =& \,G_R(x,z) + G_A(x,z),
\end{align}
where  $\vec{R} = \vec{r} - \vec{z}$. One can easily rewrite $G(x,z)$ in terms of Synge's world function $\Omega_0$ as,
\begin{equation}
G\left(x,z\right)  = \frac{1}{4\pi} \delta\left(-\Omega_0\left(x,z\right)\right)\,,
\label{green.flat}
\end{equation}
where the world function for flat spacetime is defined as \cite{J.L.Synge:1960zz},
\begin{align}
\Omega_0\left(x,z\right) &= \frac{1}{2}\eta_{\mu \nu}\left(x^{\mu} - z^{\mu}\right)\left(x^{\nu} - z^{\nu}\right) = \frac{1}{2} \left(-\left(t-z^0\right)^2 + \left(\vec{r} - \vec{z}\right)^2\right). 
\end{align}
$\Omega_0$, which is a covariant quantity measures half of the square of the proper time between two spacetime points. However $G(x,z)$ contains contributions from the advanced Green's function as well as the retarded one. Therefore to maintain causality, we can ensure that the equation 
\begin{equation}
\Box \psi(x) = - \int \limits_{-\infty}^{\infty} \delta^{4}\left(x-z(s)\right) f(s) ds
\label{flat.green}
\end{equation}
gets only contribution from the retarded Green function solution by considering the upper limit of the integral at a finite value $s_0$, 
\begin{align}
\psi(x) &=  \frac{1}{4\pi}\int \limits_{-\infty}^{s_0} \delta\left(-\Omega_0\left(x,z\right)\right) f(s) ds .
\label{sol.flat}
\end{align}
Here $z^{\mu}(s)$ is the parametric equation of the path followed by the source. In evaluating (\ref{sol.flat}), we need to choose $s_0$ such that $z^{\mu}(s_0)$ lies outside the light cone centred at $x^{\mu}$. This assures that the contribution to the scalar profile $\psi$ is only coming from the retarded part of the Green's function. One important aspect of expressing the Green's function in terms of Synge's world function $\Omega_0$ is that it is covariant by construction and hence can be used for curved spacetime as well. Thus, analogous to $\Omega_0\left(x,z\right)$ in flat spacetime, we define a world function $\Omega\left(x,z\right)$ on curved spacetime as
\begin{equation}
\Omega(x,z)=\frac{1}{2}(u_1-u_0)\int_{u_0}^{u_1}~g_{\alpha\beta}U^{\alpha}U^{\beta}~du\label{wf},
\end{equation}
where  $x$ (observer) and $z$ (source) are coordinates of two points of spacetime joined by an unique geodesic $\xi^{\alpha}$ with affine parameter `u' and $U^{\alpha}=\frac{d\xi^{\alpha}}{du}$ is the tangent to the geodesic. Gravitational radiation follows this path.

 We take following ansatz as a solution of (\ref{scalar.cov}) in terms of $\Omega\left(x,z\right)$ :
\begin{align}
\psi^{(0)}(x) &=  \frac{1}{4\pi}\int \limits_{-\infty}^{s_0} \delta\left(-\Omega\left(x,z\right)\right) f(s) ds
\label{solu.curved}.
\end{align}
When we put back $\psi^{(0)}(x)$ from (\ref{solu.curved}) into (\ref{scalar.cov}), it gives both, the source term and extra terms proportional to Riemann tensor of the background metric. To accommodate these additional terms proportional to Riemann tensor we need to find a modified solution valid up to first order in both $\phi$ and $\frac{1}{l^2}$. To do this we correct the ansatz $\psi^{(0)}(x)$ to $\psi^{(1)}(x) = \psi^{(0)}(x) + \delta\psi(x)$. In general, $\delta \psi$ is the contribution of back-scattering due to the black hole potential.
\\
~~\\
To proceed we first calculate modified Synge's world function $\Omega$ for our chosen background. Let $\xi^{\mu}(u)$ be the parametric solution of an unique geodesic between the points $x(t,\vec{x})$ and $z(z^0,\vec{z})$. To calculate the Green's function in Anti-deSitter spacetime we find out the world function $\Omega$ using its another equivalent covariant definition as \cite{J.L.Synge:1960zz},
\begin{equation}
\Omega=-\frac{1}{2}(\Delta S)^2,
\end{equation}
where
\begin{equation}
\Delta S=\int_{geodesic~ path} ds=\int_{z_0}^{t}\frac{d\xi^0}{\left(\frac{d\xi^{0}}{ds}\right)}.
\end{equation}
 To compute $\frac{d\xi^{0}}{ds}$ we use the geodesic equation for the time coordinate,
\begin{equation}
\frac{d^2\xi^{0}}{ds^2}+2~ \Gamma^0_{0k}\left(\frac{d\xi^{k}}{ds}\right)\left(\frac{d\xi^{0}}{ds}\right)=0
\end{equation}
Next to find $\left(\frac{d\xi^{0}}{ds}\right)$ we require to express $\Gamma^0_{0k}$ as a total derivative,
\begin{equation}
\frac{d^2\xi^{0}}{ds^2}+2~ \bar{\phi}_{,k}\left(\frac{d\xi^{k}}{ds}\right)\left(\frac{d\xi^{0}}{ds}\right)=0
\end{equation}
where
\begin{equation}
\bar{\phi}=\phi+\frac{\rho^2}{2l^2};\quad \phi=-\frac{M}{\rho}.
\end{equation}
This gives the solution for $\left(\frac{d\xi^{0}}{ds}\right)$ as,
\begin{equation}
\left(\frac{d\xi^{0}}{ds}\right)=A e^{-2\bar{\phi}}=A\left(1-2\bar{\phi}\right)=A\left(1-2\phi-\frac{\rho^2}{l^2}\right)\label{xis},
\end{equation}
where `$A$' is the integration constant. Therefore we can get the expression for $\Delta S$ as
\begin{equation}
\Delta S=\int_{z_0}^{t} \frac{1}{A}\left(1+2\phi+\frac{\rho^2}{l^2}\right)d\xi^0\label{S}.
\end{equation}
The constant `$A$' can be fixed by using the following property for timelike curves
\begin{equation}
g_{\mu\nu}\frac{d\xi^{\mu}}{ds}\frac{d\xi^{\nu}}{ds}=-1\label{time}.
\end{equation}
Using the metric expressions (\ref{time}) gives
\begin{equation}
-\left(1+2\phi+\frac{\rho^2}{l^2}\right)\left(\frac{d\xi^{0}}{ds}\right)^2+\delta_{ij}\left(1-2\phi+\frac{\rho^2}{2l^2}\right)\left(\frac{d\xi^{i}}{ds}\right)\left(\frac{d\xi^{j}}{ds}\right)=-1\label{radvel}
\end{equation}
Further denoting $v_i=\frac{d\xi^i}{d\xi^0}$ we can rewrite the above expression as,
\begin{equation}
\left(1+2\phi+\frac{\rho^2}{l^2}\right)-\left(1-2\phi+\frac{\rho^2}{2l^2}\right)v^2=\left(\frac{ds}{d\xi^0}\right)^2=\frac{1}{A^2}e^{4\bar{\phi}}=\frac{1}{A^2}\left(1+4\phi+\frac{2\rho^2}{l^2}\right)\label{vv}
\end{equation}
To extract `$A$' from the above expression, we again go back to the flat space limit. In $\phi \rightarrow 0$ and $\rho\rightarrow \infty$ i.e. $l\rightarrow \infty$ limit where $\rho^2/l^2\ll 1$ we define,
\begin{equation}
1-v_A^2=\frac{1}{A^2}.
\end{equation}
Here $v_A$ is the asymptotic velocity of the probe mass in flat spacetime. Keeping  terms up to linear order in `$\phi$' and ${\cal{O}}(1/l^2)$ we get from (\ref{vv}),
\begin{equation}
1-v^2=1-v_A^2-4\phi-\frac{2\rho^2}{l^2}
\end{equation}
which finally gives,
\begin{equation}
v=v_A+2\phi+\frac{\rho^2}{l^2}.
\end{equation}
The condition that the path goes through ($t, \vec{x}$) and  ($z_0, \vec{z}$) ensures
\begin{equation}
\int_{\bf{z}}^{\bf x} d\xi^i =R=\int_{z_0}^{t} v(\xi^0)~ d\xi^0=\int_{z_0}^{t} \left(v_A+2\phi+\frac{\rho^2}{l^2}\right)d\xi^0\label{R}
\end{equation}
where $R=|\vec{x}-\vec{z}|$. To carry out the integration we choose the following parametrization for $\xi^{\mu}$
\begin{equation}
\xi^{\alpha}=\xi^{\alpha}_1+s~(\xi^{\alpha}_2-\xi^{\alpha}_1)\label{para}
\end{equation}
where $\xi^0$ corresponds time and $\xi^i$ corresponds spatial coordinates. Using this parametrization, the integration now limits from $0$ to $1$. The integration involving the second and third terms of the r.h.s of (\ref{R}) are,
\begin{align}
\int_{z_0}^{t}\left(-\frac{2M}{\rho}\right)d\xi^0&=-2M\frac{(t-z_0)}{R}\log\left(\frac{\vec{x}.\vec{R}+|\vec{x}||\vec{R}|}{\vec{z}.\vec{R}+|\vec{z}||\vec{R}|}\right)\notag\\\int_{z_0}^{t}\left(\frac{\rho^2}{l^2}\right)d\xi^0&=\frac{1}{l^2}(t-z^0)\vec{x}.\vec{z}
\end{align}
where we have used the fact that $\rho^2=\xi^i\xi_i$. After carrying out the whole integration, from (\ref{R}) we get
\begin{equation}
R=~ (t-z_0)\left(v_A+\frac{1}{l^2}\vec{x}.\vec{z}-\frac{2M}{R}\log\left(\frac{\vec{x}.\vec{R}+|\vec{x}||\vec{R}|}{\vec{z}.\vec{R}+|\vec{z}||\vec{R}|}\right)\right)\label{vrA1}.
\end{equation}
Inverting the above relation, we get the asymptotic velocity $v_A$ as, 
\begin{equation}
v_A=\frac{R}{(t-z_0)}-\frac{1}{l^2}\vec{x}.\vec{z}+\frac{2M}{R}\log\left(\frac{\vec{x}.\vec{R}+|\vec{x}||\vec{R}|}{\vec{z}.\vec{R}+|\vec{z}||\vec{R}|}\right).
\end{equation} 
Finally we put this in the following integral
\begin{equation}
\Delta S=\int_{z_0}^{t} (1-v_A^2)^{1/2}\left(1+2\phi+\frac{\rho^2}{l^2}\right)d\xi^0\label{S1}
\end{equation}
and evaluate the expression for $(\Delta S)^2$ as
\begin{align}
(\Delta S)^2&=\left[(t-z_0)^2-R^2+\frac{R^2}{l^2}2\vec{x}.\vec{z}-4MR\Gamma\right]\left(1-\frac{4M}{R}\Gamma+\frac{2}{l^2}\vec{x}.\vec{z}\right)\notag\\&=\left[(t-z_0)^2-\left(\left(1-\frac{\vec{x}.\vec{z}}{l^2}\right)\left(R+2M\Gamma\right)\right)^2\right]\left(1-\frac{4M}{R}\Gamma+\frac{2}{l^2}\vec{x}.\vec{z}\right)\label{eqads}.
\end{align}
Here we denote
\begin{equation}
\Gamma\left(\vec{x},\vec{z}\right) = \text{ln} \left(\frac{\vert\vec{x}\vert \vert\vec{R}\vert + \vec{x}.\vec{R} }{\vert\vec{z}\vert\vert\vec{R}\vert + \vec{z}.\vec{R}}\right) \,.
\end{equation}
Evaluating $\delta\left(\frac{1}{2}(\Delta S)^2\right)$, 
we can write the expression of $\psi^{(0)}(x)$ from (\ref{solu.curved}),
\begin{align}
\psi^{(0)}(x)&=\frac{1}{4\pi}\int_{-\infty}^{s_0}\frac{\delta\left((t-z_0)-\left(1-\frac{\vec{x}.\vec{z}}{l^2}\right)\left(R+2M\Gamma\right)\right)}{|\vec{R}|}f(s)ds\notag\\=&\frac{1}{4\pi}\int_{-\infty}^{s_0}\frac{\delta\left((t-z_0)-R+\frac{R}{l^2}\vec{x}.\vec{z}-2M\Gamma\right)}{|\vec{R}|}f(s)ds\label{0}.
\end{align}
$\psi^0(x)$ is not a complete solution of (\ref{scalar.cov}) in presence of spacetime curvature. We identify the additional corrections due to curvature using the following property of derivative of the World function \cite{J.L.Synge:1960zz,Peters:1966}
\begin{align}
\Omega_{;\alpha}{}^{\alpha}&=4+\frac{1}{u_1-u_0}\int_{u_0}^{u_1}~(u-u_0)^2R_{\alpha\beta}U^{\alpha}U^{\beta}du+{\cal{O}}(R^2)\notag\\&=4+F(x,z)+{\cal{O}}(R^2)\label{wfd},
\end{align}
where we define $F\left(x,z\right)$ as
\begin{equation}
F\left(x,z\right) = \frac{1}{u_1 - u_0} \int \limits_{u_0}^{u_1} \left(u-u_0\right)^2 R_{\mu \nu} U^{\mu} U^{\nu} du\,\label{F}.
\end{equation}
Using (\ref{solu.curved}) and (\ref{wfd}) one can then find \cite{J.L.Synge:1960zz},
\begin{equation}
\psi^{(0)}_{\phantom{(0)};\alpha}{}^{\alpha} = - \int \limits_{-\infty}^{\infty} \delta^{4}\left(x-z(s)\right) f(s) ds + \frac{1}{4 \pi}\int \limits_{-\infty}^{s_0} \delta'\left(-\Omega\right) F(x,z(s)) f(s) ds + \mathcal{O}\left(R^2\right)
\label{psi.0},
\end{equation}
where
\begin{equation}
\delta'\left(-\Omega\right) = \frac{d \delta\left(-\Omega\right)}{d s} \frac{ds}{d\left(-\Omega\right)}\,.
\end{equation} 
Comparing the above expression with the initial one in (\ref{scalar.cov}) we get, 
\begin{equation}
\delta \psi^{(0)}_{\phantom{(0)};\alpha}{}^{\alpha} = -  \frac{1}{4 \pi}\int \limits_{-\infty}^{s_0} \delta'\left(-\Omega\right) F(x,z(s)) f(s) ds. \label{bbs}
\end{equation}
Thus the complete solution valid up to $\mathcal{O}\left(R\right)$ order is given as
\begin{align}
\psi^{(1)}(x) &= \psi^{(0)}(x) + \delta \psi^{(0)}(x) \notag\\
&= \psi^{(0)}(x) + \frac{1}{16 \pi^2} \int \sqrt{-g(y)} \delta\left(-\Omega\left((x,y\right)\right) d^4y  \int \limits_{-\infty}^{s_0} \delta'\left(-\Omega\left(y,z(s)\right)\right) F\left(y,z(s)\right) f(s) ds\,.
\label{psi1.sol}
\end{align}
Note that while $\psi^{(0)}$ involve both curvature independent and linear in curvature terms, $\delta \psi^{(0)}(x)$ only involves curvature dependent terms. The above solution of $\psi^{(1)}(x)$ satisfies 
\begin{equation}
\psi^{(1)}_{\phantom{(1)};\alpha}{}^{\alpha} = -\int \delta(x,z(s)) f(s) ds + \mathcal{O}\left(R^2\right)\,,
\label{scalar.cov2}
\end{equation}
In evaluating (\ref{psi1.sol}) the explicit expression for `F' as defined in (\ref{F}) is required.  We compute it by contracting the Einstein equation for radiation with $U^{\alpha}U^{\beta}$. Thus we get \footnote{as for null rays
	\begin{equation}
	g_{\alpha\beta}U^{\alpha}U^{\beta}=0\label{null}
	\end{equation}}
\begin{equation}
R_{\alpha\beta}U^{\alpha}U^{\beta}=8\pi GT_{\alpha\beta}U^{\alpha}U^{\beta}\label{nullein}.
\end{equation}
We recall the fact that we are considering the black hole as a point particle of mass `$M$'. Thus the stress energy tensor on curved background in (\ref{nullein}) can be expressed as,
\begin{equation}
T^{\mu\nu}(x)=M\int ds~ U^{\mu}U^{\nu}\delta^{(4)}(x-z(s))\label{T}
\end{equation} 
Using the above relations we get the correction term due to back scattering as,
\begin{align}
\delta \psi^{(0)}(x) & = \frac{GM}{2 \pi}\partial_t \int \limits_{0}^{\infty} dv \int \limits_{-\infty}^{\infty} ds \frac{\delta\left(t - z^0 - \vert \vec{z} \vert - v - \rho(v)\right)}{\rho(v) \left(\vert \vec{z}\vert + v\right)} f(s) \,, \label{dp.sol}
\end{align}
where as before $\vec{R} = \vec{x} - \vec{z}$ and 
\begin{equation}
 \rho(v) = \sqrt{\vec{x}^2 + v^2 + \frac{2 v \vec{x}.\vec{z}}{\vert \vec{z}\vert}}.
\end{equation}

Thus, we get a complete solution for the scalar box equation in AdS-Schwarzschile background with arbitrary source function. But the equation (\ref{maineom}) that we need to solve  contains extra terms proportional to the $M$ and $\frac{1}{l^2}$. To manipulate these terms, we define new ``Green's functions" as follows.
Let us first write the solution $\psi^{(1)}$ in frequency space, by performing a Fourier transform in the time coordinate as
\begin{equation}
\tilde{\psi}^{(1)} \left(\omega, \vec{x}\right) = \int dt e^{i\omega t} \psi^{(1)} \left(t, \vec{x}\right).
\end{equation}
Explicit evaluation of the Fourier transform using relations (\ref{0}) and (\ref{dp.sol}) gives,
\begin{align}
\tilde{\psi}^{(0)} & = \frac{1}{4 \pi} \int \limits_{-\infty} ^{\infty} \frac{e^{i \omega\left(z^0 + \vert \vec{R}\vert\right)}}{\vert \vec{R}\vert} f(s) ds + \frac{i \omega M}{2 \pi}\int \limits_{-\infty} ^{\infty} \frac{e^{i \omega\left(z^0 + \vert \vec{R}\vert\right)} \Gamma}{\vert \vec{R}\vert}f(s) ds\, \notag\\
&-\frac{i \omega}{4 \pi}\frac{1}{l^2}\int \limits_{-\infty} ^{\infty} e^{i \omega\left(z^0 + \vert \vec{R}\vert\right)} (\vec{x}.\vec{z})f(s) ds \label{psi0.solf}\\
\delta \tilde{\psi}^{(0)} & = -\frac{i \omega M}{2 \pi} \int \limits_{0}^{\infty} dv \int \limits_{-\infty}^{\infty} ds \frac{e^{i \omega\left(z^0 + v + \vert\vec{z}\vert + \rho(v)\right)}}{\left(v+\vert \vec{z}\vert\right)\rho(v)} f(s) \,, \label{dp.solf}
\end{align}
where for (\ref{psi0.solf}) we have expanded the exponential terms up to linear order in perturbation parameters $M$ and $1/l^2$. We further split out $\tilde{\psi}^{(0)}$ as
$
\tilde{\psi}^{(0)} = \tilde{\psi}_0^{(0)} + \tilde{\psi}_1^{(0)} + \tilde{\psi}_2^{(0)} \,,
$
where $\tilde{\psi}_0^{(0)}$ manifestly represents the zeroth order solution, while $\tilde{\psi}_1^{(0)}$ is an additional contribution due to $M$ correction and $\tilde{\psi}_2^{(0)}$ is an additional contribution due to $(1/l^2)$ correction to the solution. Thus we identify
\begin{align}
\tilde{\psi}^{(0)}_0\left(\omega,\vec{x}\right) &= \frac{1}{4 \pi} \int \limits_{-\infty} ^{\infty} \frac{e^{i \omega\left(z^0 + \vert \vec{R}\vert\right)}}{\vert \vec{R}\vert} f(s) ds \,,\label{psi00.solf}\\
 \tilde{\psi}^{(0)}_1 \left(\omega,\vec{x}\right) &= \frac{i \omega M}{2 \pi}\int \limits_{-\infty} ^{\infty} \frac{e^{i \omega\left(z^0 + \vert \vec{R}\vert\right)} \Gamma}{\vert \vec{R}\vert} f(s) ds \label{psi01.solf}\\\tilde{\psi}^{(0)}_2 \left(\omega,\vec{x}\right) &= -\frac{i \omega}{4 \pi}\frac{1}{l^2}\int \limits_{-\infty} ^{\infty} e^{i \omega\left(z^0 + \vert \vec{R}\vert\right)} (\vec{x}.\vec{z})f(s) ds
\label{psi02.solf}
\end{align}
(\ref{scalar.cov2}) can then be expanded as,
\begin{equation}\label{bla}
\Box \psi^{(1)} + 4 \phi \partial_t^2 \psi^{(1)}-\frac{\rho^2}{4l^2}\partial_t^2 \psi^{(1)}+\frac{3}{4l^2}\rho^2_k \partial_k \psi^{(1)}+\frac{3\rho^2}{4l^2}\partial_k^2 \psi^{(1)}= - \int \delta^{4}\left(x - z(s)\right) f(s) ds + \mathcal{O}\left(R^2\right).
\end{equation}
Since we are looking at the expression perturbatively, in terms that are proportional to $\phi$ and $1/l^2$ in (\ref{bla}), we replace $\psi^{(1)}$ by $\psi^{(0)}$ as
\begin{align}
 \Box \psi^{(1)} + 4 \phi \partial_t^2 \psi^{(0)}-\frac{\rho^2}{4l^2}\partial_t^2 \psi^{(0)}&+\frac{3}{4l^2}\rho^2_k \partial_k \psi^{(0)}+\frac{3\rho^2}{4l^2}\partial_k^2 \psi^{(0)}\notag\\&= - \int \delta^{4}\left(x - z(s)\right) f(s) ds + \mathcal{O}\left(R^2\right).
\label{scalar.fe}
\end{align}

Performing the Fourier transformation of the above equation we get
\begin{align}
 \tilde{\Box} \tilde{\psi}^{(1)} - 4 \omega^2\phi \tilde{\psi}^{(0)}+\frac{\rho^2}{4l^2}\omega^2 \tilde{\psi}^{(0)}&+\frac{3}{4l^2}\rho^2_k \partial_k \tilde{\psi}^{(0)}+\frac{3\rho^2}{4l^2}\partial_k^2 \tilde{\psi}^{(0)}\notag\\&= -\int e^{i\omega z^0}\delta^3(\vec{x}-\vec{z})f(s)ds + \mathcal{O}\left(R^2\right).
\label{scalar.fourfe}
\end{align}
where $\tilde{\Box}=\left(\omega^2 + \partial^2_i\right)$. Collecting terms with same coefficient as the independent perturbation parameters we get the following equations, 
\begin{align}
{\text {zeroth order} :} \quad &\tilde{\Box}\tilde{\psi}^{(0)}_0=-\int e^{i\omega z^0}\delta^3(\vec{x}-\vec{z})f(s)ds\label{box0}\\
{\text {leading order in $\phi$} :} \quad &\tilde{\Box} \tilde{\psi}^{(0)}_1+\tilde{\Box} \delta\tilde{\psi}^{(0)}- 4 \phi\omega^2 \tilde{\psi}^{(0)}=0\label{box1}\\
{\text {leading order in $\frac{1}{l^2}$} :} \quad &\tilde{\Box} \tilde{\psi}^{(0)}_2+\frac{\rho^2}{4l^2}\omega^2\tilde{\psi}^{(0)}+\frac{3}{4l^2}\rho^2_k \partial_k \tilde{\psi}^{(0)}+\frac{3}{4l^2}\rho^2 \partial_k^2 \tilde{\psi}^{(0)}=0,
\label{box2}
\end{align}
 Plugging $\tilde{\psi}^{(0)}_1$, $\delta\tilde{\psi}^{(0)}$ in (\ref{box1}), we find the following equality
\begin{equation}
-\tilde{\Box} G_M\left(\omega,\vec{x},\vec{z}\right) = \left(\omega^2 + \partial^2_i\right) G_M\left(\omega,\vec{x},\vec{z}\right) = \phi\left(\vec{x}\right)\frac{e^{i\omega \vert\vec{R}\vert}}{\vert\vec{R}\vert} \,,
\label{gf.M}
\end{equation}
where we define
\begin{equation}
G_M\left(\omega,\vec{x},\vec{z}\right) = -\frac{i G M}{2 \omega} \left(\frac{e^{i \omega \vert\vec{R}\vert} \Gamma\left(\vec{x},\vec{z}\right)}{\vert\vec{R}\vert} - \int \limits_{0}^{\infty} dv \frac{e^{i \omega\left(v + \vert\vec{z}\vert + \rho(v)\right)}}{\left(v+\vert \vec{z}\vert\right)\rho(v)} \right)\label{Gm}.
\end{equation}
Similarly plugging $\tilde{\psi}^{(0)}_2$ from (\ref{psi02.solf}) in (\ref{box2}), we find the following equality
\begin{equation}
-\tilde{\Box} G_l\left(\omega,\vec{x},\vec{z}\right)= \rho^2\left(\vec{x}\right)\frac{e^{i\omega \vert\vec{R}\vert}}{\vert\vec{R}\vert}+\frac{3}{\omega^2}\partial_k\left(\rho^2\partial_k\left(\frac{e^{i\omega R}}{R}\right)\right) \,,
\label{gf.l}
\end{equation}
where we define
\begin{equation}
G_l\left(\omega,\vec{x},\vec{z}\right) = -~\frac{4i}{\omega}~ e^{i \omega \vert\vec{R}\vert} \left(\vec{x}.\vec{z}\right)\label{Gl}.
\end{equation}
(\ref{gf.M}) and (\ref{gf.l}) enable us to express $\phi$, its derivatives and $\rho^2$ and its derivatives in the equation of motion (\ref{maineom}) in terms of $\tilde{\Box} G_M$ and $\tilde{\Box} G_l$ respectively. Since $\vec{R} = \vec{x} - \vec{z}$, the derivative $\displaystyle{\nabla_{i} = \frac{\partial}{\partial x^i} + \frac{\partial}{\partial z^i}}$ vanishes when it acts on any function of $\vec{R}$ but not on $\phi$ and $\rho$. Hence we get relations as
\begin{equation}
-\nabla_i \tilde{\Box} G_M \left(\omega,\vec{x},\vec{z}\right) = \phi_{i}\frac{e^{i\omega \vert\vec{R}\vert}}{\vert\vec{R}\vert} \,, \quad -\nabla_i \nabla_k \tilde{\Box} G_M \left(\omega,\vec{x},\vec{z}\right) = \phi_{ik} \frac{e^{i\omega \vert\vec{R}\vert}}{\vert\vec{R}\vert}\label{Dl}, 
\end{equation}
and
\begin{equation}
-\nabla_i \tilde{\Box} G_l \left(\omega,\vec{x},\vec{z}\right) = \rho^2_{i}\frac{e^{i\omega \vert\vec{R}\vert}}{\vert\vec{R}\vert} \,, \quad -\nabla_i \nabla_k \tilde{\Box} G_l \left(\omega,\vec{x},\vec{z}\right) = \rho^2_{ik} \frac{e^{i\omega \vert\vec{R}\vert}}{\vert\vec{R}\vert} \label{DG}.
\end{equation}
(\ref{Dl}) and (\ref{DG}) will be used extensively to rewrite the equation of motion (\ref{maineom}) in terms of derivatives acting on $G_{M}$ and $G_{l}$.
Thus, in this section we have developed the necessary technical tools required to deal with the case of our interest (\ref{maineom}). In the next section we present its solution.

\section{Solution of the perturbed field equations}\label{sc6}

We now look for a solution of (\ref{maineom}) for $\bar{h}_{ij}$. As it turns out, finding a solution in the frequency space is simpler. We solve the equation perturbatively to first order in parameters $M$ and 
$1/l^2$. The source strength $f(s)$ takes the form,
\begin{equation}
f(s)=16\pi G m\left(1-2\phi-\frac{\rho^2}{4l^2}\right)\frac{dz^{i}}{ds}\frac{dz^{j}}{ds}.
\end{equation}
Similar to the scalar case we will choose our trial solution for $\bar{h}_{ij}$.
At zeroth order of $M$ and $1/l^2$, the equation (\ref{maineom}) becomes,
\begin{align}
\Box\bar{h}_{ij}=-16\pi G m \int \delta^4(x,z(s))\frac{dz^{i}}{ds}\frac{dz^{j}}{ds}\,ds\label{aa}.
\end{align}
Here $\bar{h}_{ij}$ is the spatial component of trace reversed metric perturbation introduced in (\ref{trmp}), $m$ is the probe particle mass and $z^{i}$ refer to the spatial coordinate of the position of probe mass. 
The Fourier transform of the solution of (\ref{aa}) in frequency space takes the form
\begin{align}
\bar{\mathfrak{h}}_{ij}=4Gm\int dz^{0}\frac{dz^{0}}{ds} v_i v_j \frac{e^{i\omega(z^0+\vert\vec{R}\vert)}}{\vert\vec{R}\vert}\ ,\label{hijo}
\end{align}
where $\bar{\mathfrak{h}}_{ij}$ is the Fourier transform of $\bar{h}_{ij}$, $\omega$ is its frequency and $v_i=\frac{dz^i}{dz^0}$. The gauge fixing equation (\ref{gauge}) and (\ref{FE}) following a Fourier transformation yields
\begin{align}
\bar{\mathfrak{h}}_{km,m}=-i\omega \bar{\mathfrak{h}}_{k0},\quad \bar{\mathfrak{h}}_{k0,k}=-i\omega \bar{\mathfrak{h}}_{00}\label{hij00}.
\end{align}
Using these relations and equation (\ref{hijo}), we get $\bar{\mathfrak{h}}_{i0}$ and $\bar{\mathfrak{h}}_{i0}$ at zeroth order as,
\begin{align}
\bar{\mathfrak{h}}_{i0}&=-4Gm\int dz^{0}\frac{dz^{0}}{ds} v^i \frac{e^{i\omega(z^0+\vert\vec{R}\vert)}}{\vert\vec{R}\vert}\notag\\\bar{\mathfrak{h}}_{00}&=4Gm\int dz^{0}\frac{dz^{0}}{ds} \frac{e^{i\omega(z^0+\vert\vec{R}\vert)}}{\vert\vec{R}\vert}\label{hz}.
\end{align}
Finally Fourier transforming the equation (\ref{maineom}) we get,
\begin{align}
&\tilde{\Box}\left(\left(1+2\phi-\frac{\rho^2}{2l^2}\right)\bar{\mathfrak{h}}_{ij}\right)-4\left[\omega^2\phi\bar{\mathfrak{h}}_{ij}+i\omega\phi_{i}\bar{\mathfrak{h}}_{j0}+i\omega\phi_{j}\bar{\mathfrak{h}}_{i0}-\frac{1}{2}(\phi_{ij}-\frac{\delta_{ij}}{2}\phi_{kk})(\bar{\mathfrak{h}}_{00}+\bar{\mathfrak{h}}_{ll})\right]\notag\\&-\frac{1}{2l^2}\left[\omega^2\rho^2 \bar{\mathfrak{h}}_{ij}+i\omega\rho^2_{i}\bar{\mathfrak{h}}_{j0}+i\omega\rho^2_{j}\bar{\mathfrak{h}}_{i0}-\rho_{ij}^2(2\bar{\mathfrak{h}}_{00}-\bar{\mathfrak{h}}_{ll})-\frac{3}{2}\delta_{ij}\rho^2_{kl}\bar{\mathfrak{h}}_{kl}+\frac{1}{2}\delta_{ij}\rho^2_{kk}\bar{\mathfrak{h}}_{ll}-\frac{3}{2}(\rho^2_{ki}\bar{\mathfrak{h}}_{kj}+\rho^2_{kj}\bar{\mathfrak{h}}_{ki})\right.\notag\\&\left.-\rho^2_{kk}\bar{\mathfrak{h}}_{ij}-\frac{3}{2}\rho_k^2\bar{\mathfrak{h}}_{ij,k}\right]
=-16\pi G m \int ~\frac{1}{\left(1+2\phi+\frac{\rho^2}{4l^2}\right)}e^{i\omega z_0}\delta^3(\vec{x}-\vec{z}(s))\frac{dz^{i}}{ds}\frac{dz^{j}}{ds}\,ds
\label{maineomf}
\end{align}
where $\tilde{\Box}=\left(\omega^2 + \partial^2_i\right)$. Thus
plugging the expressions of the zeroth order perturbations from (\ref{hij00}) and (\ref{hz}) in source terms of (\ref{maineomf}) we get the solution for $\bar{{\mathfrak{h}}}_{ij}$ as,
\begin{align}
&\bar{\mathfrak{h}}_{ij}\left(\omega,\vec{x}\right) = \frac{4 G m}{\left(1 + 2 \phi-\frac{\rho^2}{2l^2}\right)}_{(\vec{x})} \int \frac{e^{i \omega\left(z^0 + \vert \vec{R}\vert\right)}}{\vert \vec{R}\vert} \frac{v_i v_j}{\left(1 + 2 \phi+\frac{\rho^2}{4l^2}\right)}_{(\vec{z})} \, \frac{dz^0}{ds}dz^0\notag\\&-
\int dz^0\frac{dz^0}{ds}e^{i \omega z^0} \int d^3\vec{r}~'\, \delta^{(3)} \left(\vec{r}~' - \vec{z}\right) \notag\\& \Bigg\{16 G m \left[\omega^2 v_i v_j - i \omega \left(v_i\nabla_j + v_j\nabla_i\right)-~ \frac{1}{2}\left(1+\vec{v}^2\right)\left(\nabla_i \nabla_j - \frac{1}{2}\delta_{ij}\nabla^2\right)\right]\Bigg\} G_M\left(\omega,\vec{x},\vec{r}'\right) \notag\\& -
\int dz^0\frac{dz^0}{ds}e^{i \omega z^0} \int d^3\vec{r}~'\, \delta^{(3)} \left(\vec{r}~' - \vec{z}\right) \notag\\&\Bigg\{\frac{2 G m}{l^2} \left[\omega^2 v_i v_j - i \omega \left(v_i\nabla_j + v_j\nabla_i\right)-~ \left(2-\vec{v}^2\right)\nabla_i \nabla_j - \frac{3}{2}\delta_{ij}v_k v_m\nabla_k\nabla_m- \frac{1}{2}\delta_{ij}v^2\nabla^2\right.\notag\\&\left.- \frac{3}{2}\left(v_k v_j\nabla_k\nabla_i+v_k v_i\nabla_k\nabla_j\right)+v_iv_j\nabla^2+\frac{3}{8}i\omega(v_i\nabla_j+v_j\nabla_i)\right]\Bigg\} G_l\left(\omega,\vec{x},\vec{r}'\right)\label{pr}
\end{align}
The above expression is the solution of gravitational wave front up to first order in perturbation parameters $\phi$ and $1/l^2$ and is one of the important results of this paper.  Physically $\bar{\mathfrak{h}}_{ij}\left(\omega,\vec{x}\right)$ measures the radiation, upto order ${\cal{O}}(\phi)$ and ${\cal{O}}(1/l^2)$, of frequency $\omega$ at a spatial point $\vec x$ induced by the probe particle of mass $m$ on AdS-Schwarzschild spacetime .

For the next part of the paper our aim is to consider the soft limit expansion in frequency of the above gravitational perturbation. We present the definition for soft limit in the next section. For this purpose, we follow a similar prescription as proposed in \cite{Laddha:2018myi}. To compare our results to them in appropriate limit, we also follow similar notations as in \cite{Laddha:2018myi} where the observer is assumed to be at $\vec{x}$, the trajectory of probe particle is denoted by $\vec{r}(t)$, $8\pi G=1$ and $\tilde{e}_{ij}=\bar{\mathfrak{h}}_{ij}/2$. In this notation \ref{pr} can be rewritten as,

\begin{align}
&\tilde{e}_{ij}\left(\omega,\vec{x}\right) = \frac{m}{4\pi}\frac{1}{\left(1 + 2 \phi(\vec{x})-\frac{x^2}{2l^2}\right)} \int \frac{e^{i \omega\left(t + \vert \vec{R}\vert\right)}}{\vert \vec{R}\vert} \frac{v_i v_j}{\left(1 + 2 \phi(\vec{r}(t))+\frac{r^2}{4l^2}\right)} \, \frac{dt}{ds}dt\notag\\&-
\int dt\frac{dt}{ds}e^{i \omega t} \int d^3\vec{r}~'\, \delta^{(3)} \left(\vec{r}~' - \vec{r}\right) \notag\\& \Bigg\{\frac{m}{\pi} \left[\omega^2 v_i v_j - i \omega \left(v_i\nabla_j + v_j\nabla_i\right)-~ \frac{1}{2}\left(1+\vec{v}^2\right)\left(\nabla_i \nabla_j - \frac{1}{2}\delta_{ij}\nabla^2\right)\right]\Bigg\} G_M\left(\omega,\vec{x},\vec{r}~'\right) \notag\\& -
\int dt\frac{dt}{ds}e^{i \omega t} \int d^3\vec{r}~'\, \delta^{(3)} \left(\vec{r}~'-\vec{r}\right) \notag\\&\Bigg\{\frac{m}{8\pi l^2} \left[\omega^2 v_i v_j - i \omega \left(v_i\nabla_j + v_j\nabla_i\right)-~ \left(2-\vec{v}^2\right)\nabla_i \nabla_j\right.- \frac{3}{2}\delta_{ij}v_k v_m\nabla_k\nabla_m- \frac{1}{2}\delta_{ij}v^2\nabla^2\notag\\&\left. - \frac{3}{2}\left(v_k v_j\nabla_k\nabla_i+v_k v_i\nabla_k\nabla_j\right)+v_iv_j\nabla^2+\frac{3}{8}i\omega(v_i\nabla_j+v_j\nabla_i)\right]\Bigg\} G_l\left(\omega,\vec{x},\vec{r}~'\right)
\end{align}
where $\nabla_i=\frac{\p}{\p x^i}+\frac{\p}{\p r'^i}$. In the Asymptotic limit where $\vec{x}\gg\vec{r}$, we can write, 
\begin{equation}
\vert\vec{R}\vert=\vert \vec{x}-\vec{r}\vert=R-\vec{r}.\hat{n},~~\text{where}~~ R=\vert~\vec{x}~\vert, ~\hat{n}=\frac{\vec{x}}{\vert \vec{x} \vert}
\end{equation}
In this limit the Green's functions take the following forms,
\begin{align}
G_M\left(\omega,\vec{x},\vec{r}~'\right) = i\frac{M}{16\pi \omega}   \frac{e^{i \omega (R-\hat{n}.\vec{r}~')}}{R}\left(ln\left(\frac{\vert \vec{r}~'\vert+\hat{n}.\vec{r}~'}{R}\right) + \int \limits_{\vert \vec{r}~'\vert+\hat{n}.\vec{r}~'}^{\infty} d\mathfrak{u} \frac{e^{i \omega \mathfrak{u}}}{\mathfrak{u}} \right),
\end{align}
where $\mathfrak{u}=v+\vert \vec{r}~'\vert+\hat{n}.\vec{r}~'+v \hat{n}.\hat{r}~'$ and
\begin{equation}
G_l\left(\omega,\vec{x},\vec{r}~'\right) = -\frac{4i}{\omega} \frac{e^{i \omega (R-\hat{n}.\vec{r}~')}}{R} \left(R^2~\hat{n}.\vec{r}~'\right)\label{gl}.
\end{equation}
For convenience we write $\tilde{e}_{ij}$ as,
\begin{equation}
\tilde{e}_{ij}= \tilde{e}^{(1)}_{ij}
+ \tilde{e}^{(2)}_{ij}+\tilde{e}^{(3)}_{ij}+\tilde{e}^{(4)}_{ij}++\tilde{e}^{(5)}_{ij}+\tilde{e}^{(6)}_{ij}+\tilde{e}^{(7)}_{ij}\, . \label{esmgr}.
\end{equation}
Each `$\tilde{e}$' s will correspond to different power of frequency $\omega$ in the final result. We denote the term independent of the Green's functions as $\tilde{e}^{(1)}_{ij}$,
\begin{equation} 
\tilde{e}^{(1)}_{ij}(\omega, \vec{x}) 
=\frac{m~e^{i\omega R}}{ 4\, \pi\, R}\frac{1}{\left(1-\frac{x^2}{2l^2}\right)} \int \frac{dt}{\left(1+2\Phi(\vec{r}(t))+\frac{r^2}{4l^2}\right)}\frac{dt}{ds}\, v_i v_j \, 
e^{i\omega (t-\hat{n}.\vec{r}(t))} +\text{boundary terms}
 \, ,\label{e25}
\end{equation}
where $\Phi(\vec{r})$ is the gravitational potential due to teh probe mass $M_0$,
\begin{equation} 
\Phi(\vec{r}) = -\frac{M_0}{8\pi |\vec{r}|}\, .
\end{equation}
 
Next we present the $\tilde e^{(i)}$'s which arises due to the contribution from Schwarzschild part similar to the results of \cite{Laddha:2018myi},
\begin{align}
\tilde e^{(2)}_{ij}(\omega, \vec{x}) &= i\, \frac{M_0 m}{32\pi^2\omega} \frac{e^{i\omega R}}{R}\int dt \frac{dt}{ds}\, (1+\vec{v}^2) \, \left(\n_i\n_j -\frac{1}{2} \delta_{ij} \, \n_k\n_k\right)\, \bigg\{
\ln \frac{|\vec{r}~'|+\hat{n}.\vec{r}~'}{ R} \, e^{i\omega (t - \hat{n}.\vec{r}~')} \notag\\ & + \int_{|\vec{r}~'|+\hat{n}.\vec{r}~'}^\infty \frac{d\mathfrak{u}}{\mathfrak{u}} e^{i\omega (t - \hat{n}.\vec{r}~'+\mathfrak{u})}\bigg\} 
\bigg|_{\vec{r}~'=\vec{r}(t)}\,  \label{e26} 
\end{align}
\begin{align}  \label{e28}
\tilde e^{(3)}_{ij}(\omega, \vec{x})&= -i\, \frac{M_0 m}{16\, \pi^2} \, \omega \, \frac{e^{i\omega R}}{R}\int dt \frac{dt}{ds}\, v_i\, v_j \,\bigg\{ \ln \frac{|\vec{r}~'|+\hat{n}.\vec{r}~'}{ R} \, e^{i\omega (t - \hat{n}.\vec{r}~')} \notag\\ & + \int_{|\vec{r}~'|+\hat{n}.\vec{r}~'}^\infty \frac{d\mathfrak{u}}{\mathfrak{u}} e^{i\omega (t - \hat{n}.\vec{r}~'+\mathfrak{u})}\bigg\} \, 
\end{align}
\begin{align} \label{e28aa}
\tilde e^{(4)}_{ij}(\omega, \vec{x}) =& - \frac{M_0 m}{16\, \pi^2} \frac{e^{i\omega R}}{R}\int dt \frac{dt}{ds}\, \left(v_i\n_j + v_j \n_i \right)\, 
\bigg\{ \ln \frac{|\vec{r}~'|+\hat{n}.\vec{r}~'}{ R} \, e^{i\omega (t - \hat{n}.\vec{r}~')} \notag\\ & + \int_{|\vec{r}~'|+\hat{n}.\vec{r}~'}^\infty \frac{d\mathfrak{u}}{\mathfrak{u}} e^{i\omega (t - \hat{n}.\vec{r}~'+\mathfrak{u})}\bigg\} 
\bigg|_{\vec{r}~'=\vec{r}(t)},
\end{align}
The contributions from the small cosmological constant are given by the remaining  $\tilde{e}$'s as,
\begin{align}
\tilde e^{(5)}_{ij}(\omega, \vec{x}) =& i\frac{m}{2\pi l^2 \omega}\int dt \frac{dt}{ds}\left[-~ \left(2-\vec{v}^2\right)\n_i \n_j - \frac{3}{2}\delta_{ij}v_k v_m\n_k\n_m- \frac{1}{2}\delta_{ij}v^2\n_k\n_k\right.\notag\\&\left.- ~\frac{3}{2}\left(v_k v_j\n_k\n_i+v_k v_i\n_k\n_j\right)+v_iv_j\n_k\n_k\right]\left(R~e^{i\omega (R+t - \hat{n}.\vec{r}~')}~\hat{n}.\vec{r}~'\right)\bigg|_{\vec{r}~'=\vec{r}(t)},
\end{align}
\begin{align}
\tilde e^{(6)}_{ij}(\omega, \vec{x}) =& i\frac{m}{2\pi l^2} \omega\int dt \frac{dt}{ds} v_i v_j \left(R~e^{i\omega (R+t - \hat{n}.\vec{r}~')}~\hat{n}.\vec{r}~'\right),
\end{align}
\begin{align}
\tilde e^{(7)}_{ij}(\omega, \vec{x}) =&- \frac{m}{2\pi l^2}\int dt \frac{dt}{ds} \frac{5}{8}(v_i\n_j+v_j\n_i)\left(R~e^{i\omega (R+t - \hat{n}.\vec{r}~')}~\hat{n}.\vec{r}~'\right)\bigg|_{\vec{r}~'=\vec{r}(t)}\label{e2end} ,
\end{align}
In the next section we consider soft expansion of the perturbations defined above in its frequency $\omega$. We do not discuss the expansion for $\tilde e^{(2)}_{ij}$ to $\tilde e^{(4)}_{ij}$ as it was already discussed in detail in (\cite{Laddha:2018myi}). 

\section{Soft Expansion of Radiation in asymptotically AdS space}\label{sc7}

In \cite{Laddha:2018myi}, the authors computed an interesting classical limit of the multiple soft graviton theorem, that arises while considering scattering amplitudes in a theory of gravity in asymptotically flat spacetimes. 
 To find the classical limit all operators in the soft expansion are replaced by their corresponding classical analogue. In a classical scattering, a large number of soft gravitons emerge and an important assumption is that the total energy carried by the soft radiation is small compared to the energy carried by the scatterers themselves. This is ensured by demanding a large impact parameter. While the Soft theorem is a quantum feature of a theory, in the classical limit, the soft factor gets related to the power spectrum of low frequency classical radiation that emerges during a scattering process.
In asymptotically flat backgrounds, on performing a Laurent expansion in the frequency $\omega$ of soft radiations, the classical radiative field takes the following form,
\begin{align}
&\epsilon^{\alpha\beta}\, \tilde e_{\alpha\beta}(\omega, \vec x) = {\cal{N}}'\,S_{\rm gr}(\epsilon, k) \, , \qquad {\text where}
\nonumber \\
& R \equiv |\vec x|, \quad {\cal{N}}' \equiv -\frac{i}{4\pi} {e^{i\omega R}\over R}, \quad k \equiv -\omega(1, \hat n), \quad \hat n={\vec x\over |\vec x|}\label{soft}.
\end{align}
Here $\epsilon^{\alpha\beta}$ is an arbitrary rank two polarization tensor of the graviton and $S_{\rm gr}$ is the gravitational soft factor \cite{Laddha:2018myi}.  In its soft expansion in frequency ${\omega}$, the soft factor $S_{\rm gr}$ has a term proportional to $\omega^{-1}$ in the leading order in frequency. The subleading term is proportional to $\ln\omega^{-1}$.We refer the authors to \cite{Laddha:2018rle, Laddha:2018myi, Laddha:2019yaj, Saha:2019tub} for a detailed derivation of $S_{\rm gr}$. The above equation (\ref{soft}) gives an alternate definition for the soft factor in terms of classical gravitational radiation and thus can also be computed by studying the radiation profile. 
 
For asymptotically flat backgrounds, calculation of soft factor requires consideration of large $|t|$ and suitable parametrization of $\vec r(t)$ 
where $r(t)$ is the position of the scattered mass at some time $t$. For large values of $t$ (both positive and negative), it should follow the geodesics of the background and thus the functional form can be explicitly written.
Parameterizing $\vec r(t)$ for large $|t|$ in four spacetime dimensions as in \cite{Laddha:2018myi}, we write
\begin{equation}
\vec{r}(t)=\vec{\beta}_{\pm}t-C_{\pm}\,\vec{\beta}_{\pm}\,\ln\vert t\vert+\,\text{finite terms},\qquad\vec{v}=\vec{\beta}_{\pm}\left(1-\frac{C_{\pm}}{t}\right).\label{asymrv}
\end{equation}
Here we keep terms of the order of ${\cal{O}}(1/t)$. The $\ln\vert t\vert$ terms arises due to the presence of long range interaction force in 4 spacetime dimensions. Following the parametrization (\ref{asymrv}) thereafter one can use suitable integrals deduced in \cite{Laddha:2018myi} and find the soft factors following the relation (\ref{soft}).

 Consideration of soft limit is tricky for asymptotic AdS spacetime. There is no notion of null infinity in asymptotic AdS spacetimes. Gravitons are bounced off an infinite number of times on spatial infinity, which is a time-like hypersurface. The frequency of the graviton can never strictly go to zero, as there is a mass gap. Thus, we need to suitably define a "soft limit" in this case. We take physical insight to do so and consider a double scaling limit. We consider that the frequency of the graviton in AdS space goes to zero while the radius of the AdS space goes to infinity, i.e. the AdS space limits to a flat space. Henceforth we define the soft limit as two simultaneous limits, $\omega\rightarrow 0$ and $l\rightarrow \infty$, keeping $\omega l$ fixed. 
 
 Next to understand how the long range gravitational force behave for AdS case in four spacetime dimensions and we consider Gauss’s law in AdS spcetime \cite{Kaplan} for this purpose. Let us perform a coordinate transformation on asymptotic static AdS metric to global coordinates to get 
\begin{equation}
d\tilde{s}^2= -\left(1+\frac{r^2}{l^2}\right)^2~dt^2+dr^2+r^2\left(1+\frac{r^2}{l^2}\right)d\Omega^2.
\end{equation} 
We consider a constant time surface, such as the surface t = 0 in this background. We have a Mass M sitting at the centre of the black hole r=0. Gauss’s law implies that the total amount of flux through any sphere must be a constant. The potential due to mass is
\begin{align}
V=\frac{M}{r\sqrt{1+\frac{r^2}{l^2}}}\label{V}.
\end{align}
`$r$' follows from the metric as,
\begin{align}
& dr=\sqrt{\frac{-g_{00}}{g_{rr}}}~dt=\left(1+\frac{r^2}{l^2}\right) dt\label{arb}.
\end{align}
Integrating (\ref{arb}) and considering the large $l$ limit we get,
\begin{align}
r\sim \beta l\tan\left(\frac{t}{l}\right)\Rightarrow r \sim \beta t+\frac{\beta t^3}{3 l^2} \quad \text{for large}\quad l\, .\label{rt}
\end{align}
We now replace $r$ in terms of $t$ from the above relation in the expression of potential $V$ in (\ref{V}) and we get
\begin{equation}
V \sim \frac{1}{\beta t}-\frac{1}{l^2}\left(\frac{t}{3\beta}+\frac{t\beta}{2}\right)\label{ab}.
\end{equation}
Therefore integrating (\ref{ab})\footnote{Force$= \frac{d^2 r}{dt^2}=-\frac{dV}{dr}$} we get the additional contribution in $r$ due to the long range attractive force caused by other particles involved in the scattering, as proportional to 
\begin{equation}
r \sim c_1 \ln t-\frac{c_2}{l^2}t^2,\label{tlr}.
\end{equation}
In (\ref{tlr}) the first term is same as for asymptotically flat backgrounds and the second term is the effect of the cosmological constant. Our calculations are valid only up to the order of $\frac{1}{l^2}$ and when $\frac{r^2}{l^2}\ll 1$. As the perturbations are already at liner order in $\frac{1}{l^2}$, the contribution from cosmological constant in $r$ ( second term in (\ref{tlr})) does not affect our calculation. We assume $\omega l\rightarrow \gamma$ which is a large finite number (as $\frac{r^2}{l^2}\ll 1$) in our soft limit. Thus we rewrite the final expressions of perturbations by pulling out a factor of $\frac{1}{\omega^2l^2}$  and study the dependence of the remaining terms on $\omega$ in $\omega\rightarrow 0$ limit.

To evaluate the soft expansion of $\tilde e^{(1)}_{ij}$, $\tilde e^{(5)}_{ij}$, $\tilde e^{(6)}_{ij}$ and $\tilde e^{(7)}_{ij}$ we need to find the asymptotic expression for $\frac{dt}{ds}$. It is given as, 
\begin{align}\label{edefdtds}
\frac{dt}{ds}=& =(-g_{\mu\nu}v^{\mu}v^{\nu})^{-1/2}=\left\{\left(1- \frac{M_0}{4\pi |\vec{r}(t)|}+\frac{r^2}{l^2}\right) - \left(1- \frac{M_0}{4\pi |\vec{r}(t)|}+\frac{r^2}{2l^2}\right)^{-1} \vec{v}(t)^2\right\}^{-1/2} \notag\\ &\simeq \frac{1}{\sqrt{1-\vec{v}(t)^2}} \, \left\{ 1 + \frac{M_0}{8\, \pi |\vec{r}(t)|} \frac{1+\vec{v}(t)^2}{1-\vec{v}(t)^2}-\frac{r^2}{4l^2}\frac{2-\vec{v}(t)^2}{1-\vec{v}(t)^2}\right\} \quad \text{for large} |\vec r(t)|\, .
\end{align}
We consider the expansion of $\tilde e^{(1)}_{ij}$ in details below.For that we write $e^{i\omega(t-\hat{n}.\vec{r}(t))}$ as
\begin{equation}
e^{i\omega(t-\hat{n}.\vec{r}(t))}=\frac{1}{i\omega}\frac{1}{\p_t(t-\hat{n}.\vec{r}(t))}\frac{d}{dt}e^{i\omega(t-\hat{n}.\vec{r}(t))}=\frac{1}{i\omega}\frac{1}{(1-\hat{n}.\vec{v}(t))}\frac{d}{dt}e^{i\omega(t-\hat{n}.\vec{r}(t))}\label{eas}
\end{equation} 
Thereafter carrying out the integration by parts we get
\begin{align}
&\tilde e^{(1)}_{ij}(\omega, \vec x)  = -\frac{m}{4\pi\, R}\frac{1}{\left(1-\frac{x^2}{2l^2}\right)} e^{i\omega R} \frac{1}{i\omega} \int dt \, e^{i\omega(t - \hat n. \vec r(t))} 
\, \frac{d}{dt}\left[\frac{1}{1 - \hat n.\vec v(t)} \frac{1}{1+2\Phi(\vec r(t))+\frac{r^2}{4l^2}} \, \frac{dt}{ds}\, v_i v_j \right] \notag\\=& -\frac{m}{4\pi\, R}\frac{1}{\left(1-\frac{x^2}{2l^2}\right)} e^{i\omega R} \frac{1}{i\omega} \int dt \, e^{i\omega(t - \hat n. \vec r(t))} 
\, \frac{d}{dt}\left[\frac{1}{1 - \hat{n}.\vec{v}(t)}\frac{1}{1-M_0/(4\pi |\vec r(t)|)+\frac{r^2}{4l^2}} \, \frac{dt}{ds} \, v_i v_j\right]
\end{align}
Plugging the expression of $dt/ds$ from (\ref{edefdtds}) we get, 
\begin{align}
&\tilde e^{(1)}_{ij}(\omega, \vec x) =\notag\\& 
-\frac{m}{4\pi\, R}\frac{1}{\left(1-\frac{x^2}{2l^2}\right)} e^{i\omega R} \frac{1}{i\omega} \int dt \, e^{i\omega(t - \hat n. \vec r(t))} 
\, \frac{d}{dt}\left[\frac{1}{1 - \hat{n}.\vec{v}(t)}\frac{1}{\sqrt{1-\vec{v}(t)^2}} \left\{ 1 + \frac{M_0}{8\, \pi |\vec{r}(t)|} \frac{3-\vec{v}(t)^2}{1-\vec{v}(t)^2}\right\}v_i v_j\right]\notag\\&+\frac{m}{16\pi l^2 R} e^{i\omega R} \frac{1}{i\omega}\int dt \, e^{i\omega(t - \hat n. \vec r(t))} 
\,  \frac{d}{dt}\left[\frac{1}{1 - \hat{n}.\vec{v}(t)}\frac{1}{\sqrt{1-\vec{v}(t)^2}}
\left\{r^2\frac{3-2\vec{v}(t)^2}{1-\vec{v}(t)^2}\right\} v_i v_j\right]\label{adse}
\end{align}
We denote the first line in the rhs of of the relation (\ref{adse}) by $X_1$ and the second line by $X_2$. We find that part of $X_2$ integral contributes in the soft factor and it is arising due to the presence of cosmological constant. As $t\to\pm\infty$, $X_1$ is the same contribution as the Schwarzschild case of \cite{Laddha:2018myi} multiplied by $\frac{1}{\left(1-\frac{x^2}{2l^2}\right)}$ factor. We manipulate part of $X_1$ using the expressions of $\vec{v}(t)$ given in (\ref{asymrv}) as,
\begin{align}
&\frac{1}{1 - \hat{n}.\vec{v}(t)}\frac{1}{\sqrt{1-\vec{v}(t)^2}} \left\{ 1 + \frac{M_0}{8\, \pi |\vec{r}(t)|} \frac{3-\vec{v}(t)^2}{1-\vec{v}(t)^2}\right\}v_i v_j\notag\\& =\frac{1}{ 1 - \hat{n}.\vec{\beta}_\pm} \, \frac{1}{\sqrt{1-\vec{\beta}_\pm^2}}\, \beta_{\pm\, i}~\beta_{\pm\, j}\left[ 1  - \frac{1}{t} \left\{C_\pm \frac{1}{1-\hat n.\vec{\beta}_\pm} \mp 
\frac{ M_0}{8\, \pi\, |\vec \beta_\pm|}\, \frac{3-\vec{\beta}_\pm^2}{ 1-\vec{\beta}_\pm^2} 
+C_\pm \frac{1}{1 -\vec{\beta}_\pm^2} 
\right\}\right]\, \label{nnb}.
\end{align}
Plugging back (\ref{nnb}) into the expression of $X_1$ we compare it with the following known integral $I_1$ \cite{Laddha:2018myi}, 
\begin{align}
I_1\equiv &{1\over \omega} \int_{-\infty}^\infty dt \, e^{-i\,\omega\, g(t) } f'(t)=\left[\frac{1}{\omega}(f_{+}-f_{-})+i(a_+ k_{+}-a_- k_{-})\ln \omega^{-1}\right]+\text{finite terms},\label{i1}
\end{align}
where comparing with (\ref{adse})
\begin{align}
f_\pm =& \frac{i}{\left(1-\frac{x^2}{2l^2}\right)}\, \frac{m}{4\pi\, R} e^{i\omega R}  
\frac{1}{1 - \hat{n}.\vec{\beta}_\pm} \, \frac{1}{\sqrt{1-\vec \beta_\pm^2}}\, \beta_{\pm i} 
\beta_{\pm j},\notag\\k_\pm &=-\frac{i}{\left(1-\frac{x^2}{2l^2}\right)}\,\frac{m}{4\pi\, R} e^{i\omega R}  
\frac{1}{1 - \hat{n}.\vec{\beta}_\pm} \, \frac{1}{\sqrt{1-\vec{\beta}_\pm^2}}\, \beta_{\pm i}\beta_{\pm j} \notag\\&\left\{C_\pm \frac{1}{1-\hat{n}.\vec{\beta}_\pm} \mp 
\frac{ M_0}{8\, \pi\, |\vec \beta_\pm|}\, \frac{3-\vec\beta_\pm^2}{1-\vec\beta_\pm^2} 
 +C_\pm \frac{1}{1 -\vec\beta_\pm^2}\right\}, \notag \\ 
a_\pm &= - (1 -\hat n.\vec \beta_\pm) \, .
\end{align}
Evaluating the integral and keeping terms up to ${\cal{O}}(M)$ and ${\cal{O}}(1/l^2)$ we get the expression for $X_1$ as, 
\begin{align}
X_1 &= i\,\omega^{-1}\frac{m}{4\pi\, R} e^{i\omega R} 
\left\{ {1\over 1 - \hat n.\vec \beta_+} \, {1\over \sqrt{1-\vec \beta_+^2}}\, \beta_{+ i} ~
\beta_{+ j}- {1\over 1 - \hat n.\vec \beta_-} \, {1\over \sqrt{1-\vec \beta_-^2}}\, \beta_{- i}~ 
\beta_{- j}\right\}\notag\\
& -{m\over 4\pi\, R} e^{i\omega R}  \ln \omega^{-1} \left[ 
{1\over \sqrt{1-\vec \beta_+^2}}\, \beta_{+ i} 
\beta_{+ j} \left\{C_+ {1\over 1-\hat n.\vec\beta_+} - 
{ M_0\over 8\, \pi\, |\vec\beta_+|} \, {3-\vec\beta_+^2\over 1-\vec\beta_+^2}
+C_+ {1\over 1 -\vec\beta_+^2} 
\right\} \right.\nonumber \\ &
\left. -~ {1\over \sqrt{1-\vec \beta_-^2}}\, \beta_{- i}~ 
\beta_{- j}~\left\{C_- {1\over 1-\hat n.\vec\beta_-} + 
{M_0\over 8\pi ~~|\,\vec\beta_-|} \, {3-\vec\beta_-^2\over 1-\vec\beta_-^2}
+C_- {1\over 1 -\vec\beta_-^{~2}} 
\right\}
\right]\,\notag\\&+ i\frac{x^2}{2l^2}\,\omega^{-1}\frac{m}{4\pi\, R} e^{i\omega R} 
\left\{ {1\over 1 - \hat n.\vec \beta_+} \, {1\over \sqrt{1-\vec \beta_+^2}}\, \beta_{+ i} ~
\beta_{+ j}- {1\over 1 - \hat n.\vec \beta_-} \, {1\over \sqrt{1-\vec \beta_-^2}}\, \beta_{- i}~ 
\beta_{- j}\right\}\notag\\
& + X_1^l . \label{X11}
\end{align}
The first three lines of (\ref{X11}) represents the results that arises due to the Schwarzschild background as discussed in \cite{Laddha:2018myi}. The new term $X_1^l$ in (\ref{X11}) results due to the cosmological constant and is given as,  
\begin{align}
X_1^l &=i\frac{x^2}{2\omega^2l^2}\,\frac{m}{4\pi\, R} e^{i\omega R} 
\left\{ \omega{1\over 1 - \hat n.\vec \beta_+} \, {1\over \sqrt{1-\vec \beta_+^2}}\, \beta_{+ i} ~
\beta_{+ j}- \omega{1\over 1 - \hat n.\vec \beta_-} \, {1\over \sqrt{1-\vec \beta_-^2}}\, \beta_{- i}~ 
\beta_{- j}\right\}\notag\\
& -\frac{x^2}{2\omega^2 l^2} {m\over 4\pi\, R} e^{i\omega R} \left\{\omega^2 \ln \omega^{-1} \left[ 
{1\over \sqrt{1-\vec \beta_+^2}}\, \beta_{+ i} 
\beta_{+ j} \left\{C_+ {1\over 1-\hat n.\vec\beta_+} +C_+ {1\over 1 -\vec\beta_+^2} 
\right\} \right.\right.\nonumber \\ &\left.
\left. -~ {1\over \sqrt{1-\vec \beta_-^2}}\, \beta_{- i}~ 
\beta_{- j}~\left\{C_- {1\over 1-\hat n.\vec\beta_-} +C_- {1\over 1 -\vec\beta_-^{~2}} 
\right\}\right]\right\}.\label{1sa}
\end{align}
Since $\omega l $ in constant, one can easily notice that the terms in (\ref{1sa}) are finite in $\omega\rightarrow 0$ limit and hence will not contribute to the soft factor. Next we evaluate $X_2$ from (\ref{adse}). 
By putting the expressions of $\vec{r}$ and $\vec{v}$ from (\ref{asymrv}), the total derivative term in the integrand of $X_2$ evaluates to,
\begin{align}
\frac{1}{1- \hat{n}.\vec{\beta}}\beta_{\pm i} \beta_{\pm j}\frac{1}{\sqrt{1-\vec{\beta}_\pm^2}}\, \left[2 A_1 t+ A_2 \ln t+A_2+ A_3+A_4 \frac{1}{t}+\frac{d}{dt}\left(A_5+ A_6\frac{1}{t}\right)\right], \label{A}
\end{align}
where
\begin{align}
A_1=&\frac{\vec{\beta}_{\pm}^2 (3-2\vec{\beta}_{\pm}^2)}{(1-\vec{\beta}_{\pm}^2)}, \quad A_2=-2C_{\pm}\vec{\beta}_{\pm}^2 \frac{(3-5\vec{\beta}_{\pm}^2 + 2\vec{\beta}_{\pm}^4)}{(1-\vec{\beta}_{\pm}^2)^{2}},\notag\\A_3=&\frac{C_{\pm}\vec{\beta}_{\pm}^2}{(1-\vec{\beta}_{\pm}^2)^{2}} \{-3(2+\alpha)+ 5(1+\alpha) \vec{\beta}_{\pm}^2 -2(1+\alpha) \vec{\beta}_{\pm}^4 \},\notag\\ A_4=&-\frac{1}{\left(1-\vec{\beta}_{\pm}^2\right)^{2}}\vec{\beta}_{\pm}^2 C^2_{\pm} \left(4 (\alpha +1) \vec{\beta}_{\pm}^4-10 (\alpha +1)\vec{\beta}_{\pm}^2+6 (\alpha +2)\right),\notag\\A_5=&-\frac{1}{2(1-\vec{\beta}_{\pm}^2)^{3}}C_{\pm}^2\vec{\beta}_{\pm}^2\left[-6(1+\alpha)^2+\vec{\beta}_{\pm}^2(-9+2\alpha(11+8\alpha))-14\alpha\vec{\beta}_{\pm}^2(1+\alpha)+4\alpha\vec{\beta}_{\pm}^6(1+\alpha)\right]\notag\\=& -\frac{1}{2(1-\vec{\beta}_{\pm}^2)^{3}}C_{\pm}^2\vec{\beta}_{\pm}^2\left[-6(1+\alpha)^2-9\vec{\beta}_{\pm}^2+8\alpha\vec{\beta}_{\pm}^2+2\alpha^2\vec{\beta}_{\pm}^2+4\alpha\vec{\beta}_{\pm}^6(1+\alpha)\right],\notag\\ A_6=& -\frac{1}{2(1-\vec{\beta}_{\pm}^2)^{4}}C_{\pm}^3\beta_{\pm}^2\left[2\alpha^3(-1+\vec{\beta}^2_{\pm})^3(-3+2\vec{\beta}_{\pm}^2)+5\vec{\beta}_{\pm}^2(4+3\vec{\beta}_{\pm}^2)+3\alpha(2+\vec{\beta}_{\pm}^2-3\vec{\beta}_{\pm}^4)\right.\notag\\&\left.+2\alpha^2(-1+\vec{\beta}_{\pm}^2)^2(6-5\vec{\beta}_{\pm}^2+2\vec{\beta}_{\pm}^4)\right],
\end{align}
and
\begin{equation}
\alpha=\frac{\hat{n}.\vec{\beta}_{\pm}}{1-\hat{n}.\vec{\beta}_{\pm}}.
\end{equation}
Let us first look into the part of (\ref{A}) proportional to $\frac{d}{dt}\left[A_5 +A_6\frac{1}{t}\right]$, which is the constant piece and inversely proportional to t and denote as ${\cal{I}}_1$. Comparing with \ref{i1} we can write
\begin{align}
{\cal{I}}_1&=\frac{1}{l^2}\left[\frac{1}{\omega}(f^{(1)}_{+}-f^{(1)}_{-})+i(a_+ k^{(1)}_{+}-a_- k^{(1)}_{-})\ln \omega^{-1}\right]\notag\\&=\frac{1}{\omega^2 l^2}\left[\omega(f^{(1)}_{+}-f^{(1)}_{-})+i(a_+ k^{(1)}_{+}-a_- k^{(1)}_{-})\omega^2\ln \omega^{-1}\right]\label{SN}
\end{align}
where
\begin{align}
f^{(1)}_{\pm}&=-i\frac{m}{16\pi R} e^{i\omega R} \, \frac{1}{1- \hat{n}.\vec{\beta}_{\pm}}\beta_{\pm i} \beta_{\pm j}\frac{1}{\sqrt{1-\vec{\beta}_\pm^2}}\,A_5 \notag\\k^{(1)}_{\pm}&= -i\frac{m}{16\pi R} e^{i\omega R} \, \frac{1}{1- \hat{n}.\vec{\beta}_{\pm}}\beta_{\pm i} \beta_{\pm j}\frac{1}{\sqrt{1-\vec{\beta}_\pm^2}}\,A_6,\notag\\a_\pm &= - (1 -\hat n.\vec \beta_\pm) \, .
\end{align}
We again observe that in \ref{SN} the terms are finite in $\omega\rightarrow 0$ limit and therefore do not contribute to the soft factor.
Next we denote the remaining parts of (\ref{SN}) as ${\cal{I}}_2$ that is proportional to
\begin{align*} 
 \left[ 2 A_1 t+ A_2 \ln t+A_2+ A_3 +A_4\frac{1}{t}\right]\label{dd}.
\end{align*}
Following (\ref{eas}), we manipulate the above terms as,
\begin{align}
{\cal{I}}_2=&\frac{m}{16\pi l^2 R} e^{i\omega R} \frac{1}{\omega^2}\int dt \, e^{i\omega(t - \hat n. \vec r(t))}\frac{1}{1- \hat{n}.\vec{\beta}}\beta_{\pm i} \beta_{\pm j}\frac{1}{\sqrt{1-\vec{\beta}_\pm^2}}\,\notag\\&\frac{d}{dt} \left[ \frac{1}{1 - \hat{n}.\vec{v}(t)}\left(2 A_1 t+ A_2 \ln t+A_2+ A_3 +A_4\frac{1}{t}\right)\right]
\end{align}
An integration by parts results in,
\begin{align}
{\cal{I}}_2&=\frac{m}{16\pi l^2 R} e^{i\omega R} \frac{1}{\omega^2}\int dt \, e^{i\omega(t - \hat n. \vec r(t))}\frac{\beta_{\pm i} \beta_{\pm j}}{(1- \hat{n}.\vec{\beta})^2}\frac{1}{\sqrt{1-\vec{\beta}_\pm^2}}\,\frac{d}{dt} \left[2 A_1 t-2\alpha C_{\pm}A_1+ \right.\notag\\&\left.(A_2+A_3)+A_2 \ln(t)+\frac{1}{t}\left(2\alpha^2C_{\pm}^2 A_1-\alpha C_{\pm}(A_2+ A_3) +A_4\right)-\frac{\alpha}{t}C_{\pm} A_2\ln(t)\right]\label{i22}
\end{align}
Integration for the terms of ${\cal{I}}_2$ which are constant and ${\cal{O}}(t^{-1})$ can be easily done using the integral (\ref{i1}). We denote these terms as ${\cal{I}}_2^1$ in the following and study whether they have any contribution to the soft factor.
\begin{align}
{\cal{I}}_2^1=&\frac{m}{16\pi l^2 R} e^{i\omega R} \frac{1}{\omega^2}\int dt \, e^{i\omega(t - \hat n. \vec r(t))}\frac{\beta_{\pm i} \beta_{\pm j}}{(1- \hat{n}.\vec{\beta})^2}\frac{1}{\sqrt{1-\vec{\beta}_\pm^2}}\, \notag\\&\frac{d}{dt}\left[-2\alpha C_{\pm}A_1+ (A_2+A_3)+\frac{1}{t}\left(2\alpha^2C_{\pm}^2 A_1-\alpha C_{\pm}(A_2+ A_3) +A_4\right)\right]\notag\\=&\frac{1}{\omega^2 l^2}\left[(f^{(2)}_{+}-f^{(2)}_{-})+i(a_+ k^{(2)}_{+}-a_- k^{(2)}_{-})\omega\ln \omega^{-1}\right],\label{11}
\end{align}
where
\begin{align}
f^{(2)}_{\pm}&=\frac{m}{16\pi R} e^{i\omega R}\, \frac{\beta_{\pm i} \beta_{\pm j}}{(1- \hat{n}.\vec{\beta})^2}\frac{1}{\sqrt{1-\vec{\beta}_\pm^2}}\left[-2\alpha C_{\pm}A_1+ (A_2+A_3)\right]\notag\\k^{(2)}_{\pm}&=\frac{m}{16\pi R} e^{i\omega R}\, \frac{\beta_{\pm i} \beta_{\pm j}}{(1- \hat{n}.\vec{\beta})^2}\frac{1}{\sqrt{1-\vec{\beta}_\pm^2}}\left(2\alpha^2C_{\pm}^2 A_1-\alpha C_{\pm}(A_2+ A_3) +A_4\right)
\end{align}
The terms in (\ref{11}) are finite in $\omega\rightarrow 0$ limit and hence do not contribute to the soft factor.Finally we manipulate the remaining terms in (\ref{i22}). Denoting these as ${\cal{I}}_2^2$ we get,
\begin{align}
{\cal{I}}_2^2=&\frac{m}{16\pi l^2 R} e^{i\omega R} \frac{1}{\omega^2}\int dt \, e^{i\omega(t - \hat n. \vec r(t))}\frac{\beta_{\pm i} \beta_{\pm j}}{(1- \hat{n}.\vec{\beta})^2}\frac{1}{\sqrt{1-\vec{\beta}_\pm^2}}\,\frac{d}{dt} \left[2A_1 t+A_2\ln(t)\right]\notag\\=&\frac{m}{16\pi l^2 R} e^{i\omega R} \frac{1}{\omega^2}\int dt \, e^{i\omega(t - \hat n. \vec r(t))}\frac{\beta_{\pm i} \beta_{\pm j}}{(1- \hat{n}.\vec{\beta})^2}\frac{1}{\sqrt{1-\vec{\beta}_\pm^2}}\,\left[2A_1+\frac{A_2}{t}\right].
\end{align}
Carrying out the similar by parts treatment stated earlier we get,
\begin{align}
{\cal{I}}_2^2=&i\frac{m}{16\pi l^2 R} e^{i\omega R} \frac{1}{\omega^3}\int dt \, e^{i\omega(t - \hat n. \vec r(t))}\frac{\beta_{\pm i} \beta_{\pm j}}{(1- \hat{n}.\vec{\beta})^3}\frac{1}{\sqrt{1-\vec{\beta}_\pm^2}}\,\frac{d}{dt} \left[2A_1 +\frac{1}{t}\left(A_2-2A_1C_{\pm} \alpha\right)\right]\notag\\=&\frac{1}{\omega^2 l^2}\left[\frac{1}{\omega}(f^{(3)}_{+}-f^{(3)}_{-})+i(a_+ k^{(3)}_{+}-a_- k^{(3)}_{-})\ln \omega^{-1}\right]\label{adssoft},
\end{align}
where
\begin{align}
f^{(3)}_{\pm}&=i\frac{m}{8\pi R} e^{i\omega R} \frac{\beta_{\pm i} \beta_{\pm j}}{(1- \hat{n}.\vec{\beta}_{\pm})^3}\frac{1}{\sqrt{1-\vec{\beta}_\pm^2}} A_1\notag\\k^{(3)}_{\pm}&=i\frac{m}{16\pi R} e^{i\omega R} \frac{\beta_{\pm i} \beta_{\pm j}}{(1- \hat{n}.\vec{\beta}_{\pm})^3}\frac{1}{\sqrt{1-\vec{\beta}_\pm^2}}\left(A_2-2A_1C_{\pm} \alpha\right).
\end{align}
In (\ref{adssoft}) both terms in the parenthesis diverges in $\omega\rightarrow 0$ limit and are the main contribution to the soft factor arising due to the consideration of the AdS background. In the appendix we will discuss how $\tilde{e}^{(5)}_{ij}$, $\tilde{e}^{(6)}_{ij}$ and $\tilde{e}^{(7)}_{ij}$ are finite in $\omega\rightarrow 0$ limit keeping $\omega l$ fixed and therefore they do not contribute in the soft factor.

Our final task is to compute the soft factor using the previous results. To do so we will first contract all the perturbations $\tilde e_{ij}$ with the polarization tensor $\epsilon^{ij}$ and use the relation (\ref{soft}) to predict the soft factor. The invariance of soft factor under the transformation $\epsilon^{\mu\nu}\rightarrow \epsilon^{\mu\nu}+\xi^{\mu}k^{\nu}+\xi^{\nu}k^{\mu}$ for arbitrary $\xi^{\alpha}$ and our gauge choice allows us to choose the polarization tensor as
\begin{equation}
\epsilon^{0\nu}=0,~~\epsilon^i{}_i=0; ~~k_i\epsilon^{ij}=0, 
\end{equation} 
The resulting soft factor have two parts. One of them arises due to the consideration of Schwarzschild background and the other part involves the contribution from the cosmological constant. The former part is already discussed in detail in \cite{Laddha:2018myi} which can be obtained from the first three lines of \ref{X11} and from the combination of $\tilde e^{(2)}_{ij}$ and $\tilde e^{(3)}_{ij}$. (\ref{adssoft}) produces the latter part. Finally we can write the complete soft factor as $\tilde S_{\rm gr}= \tilde S_{\rm gr}^M+\tilde S_{\rm gr}^l$, where,
\begin{eqnarray} \label{sfs}
\tilde S_{\rm gr}^M &=& -m\, \omega^{-1}\, \epsilon^{ij}\, 
\left\{ {1\over 1 - \hat n.\vec \beta_+} \, {1\over \sqrt{1-\vec \beta_+^2}}\, \beta_{+ i} 
\beta_{+ j} - {1\over 1 - \hat n.\vec \beta_-} \, {1\over \sqrt{1-\vec \beta_-^2}}\, \beta_{- i} 
\beta_{- j}\right\}\nonumber \\
&& \hskip -.5in - i\, {m} \,   \ln \omega^{-1} \, \epsilon^{ij}\, \left[ 
{1\over \sqrt{1-\vec \beta_+^2}}\, \beta_{+ i} 
\beta_{+ j} \left\{C_+ {1\over 1-\hat n.\vec\beta_+} -
{ M_0\over 8\, \pi\, |\vec\beta_+|^3}\, {3\vec\beta_+^2 -1 \over 1-\vec \beta_+^2}
+C_+ {1\over 1 -\vec\beta_+^2} 
\right\} \right.\nonumber \\ &&
\left. - {1\over \sqrt{1-\vec \beta_-^2}}\, \beta_{- i} 
\beta_{- j} \left\{C_- {1\over 1-\hat n.\vec\beta_-} + 
{M_0\over 8\, \pi\, |\vec\beta_-|^3} {3\vec\beta_-^2-1\over 1-\vec\beta_-^2}
+C_- {1\over 1 -\vec\beta_-^2} 
\right\}
\right]\, \label{Scpart}
\end{eqnarray}
and
\begin{align}
&\tilde S_{\rm gr}^l=\notag\\&-\frac{m}{2\gamma^2}\, \omega^{-1}\, \epsilon^{ij} \left\{\frac{1}{(1- \hat{n}.\vec{\beta})^3}\beta_{+ i} \beta_{+ j}\frac{1}{\sqrt{1-\vec{\beta}_+^2}} \frac{\vec{\beta}_{+}^2 (3-2\vec{\beta}_{+}^2)}{(1-\vec{\beta}_{+}^2)}-\frac{1}{(1- \hat{n}.\vec{\beta})^3}\beta_{- i} \beta_{- j}\frac{1}{\sqrt{1-\vec{\beta}_-^2}}\frac{\vec{\beta}_{-}^2 (3-2\vec{\beta}_{-}^2)}{(1-\vec{\beta}_{-}^2)}\right\}\notag\\& - i\, \frac{m}{4\gamma^2} \,   \ln \omega^{-1} \, \epsilon^{ij}\, \left[\frac{\beta_{+i} \beta_{+j}}{(1- \hat{n}.\vec{\beta}_{+})^2}\frac{1}{\sqrt{1-\vec{\beta}_+^2}}\frac{2C_+ \vec{\beta}^2_+}{(1-\vec{\beta}_+^2)^2}(3-5\vec{\beta}_{+}^2 + 2\vec{\beta}_{+}^4)\right.\notag\\&\left.+\frac{\beta_{+ i} \beta_{+ j}}{(1- \hat{n}.\vec{\beta}_+)^3}\frac{1}{\sqrt{1-\vec{\beta}_+^2}}\frac{2C_+ \hat{n}.\vec{\beta}_+}{1-\vec{\beta}_+^2}\vec{\beta}_+^2(3-2\vec{\beta}_+^2)-\frac{\beta_{-i} \beta_{-j}}{(1- \hat{n}.\vec{\beta}_{-})^2}\frac{1}{\sqrt{1-\vec{\beta}_-^2}}\frac{2C_- \vec{\beta}^2_-}{(1-\vec{\beta}_-^2)^2}(3-5\vec{\beta}_{-}^2 + 2\vec{\beta}_{-}^4)\right.\notag\\&\left.-\frac{\beta_{- i} \beta_{- j}}{(1- \hat{n}.\vec{\beta}_-)^3}\frac{1}{\sqrt{1-\vec{\beta}_-^2}}\frac{2C_- \hat{n}.\vec{\beta}_-}{1-\vec{\beta}_-^2}\vec{\beta}_-^2(3-2\vec{\beta}_-^2)\right]\label{fr}
\end{align}
The term $\tilde S_{\rm gr}^l$ proportional to $\frac{1}{\gamma^2}$ are the contributions of the cosmological constant to the soft factor. We will fix $C_{\pm}$ from the energy conservation. The total energy of the particle is given by,
\begin{equation}
E=m |g_{00}|\frac{dt}{ds}=m\left(1-\frac{M_0}{4\pi r}+\frac{r^2}{l^2}\right)\left[\left(1-\frac{M_0}{4\pi r}+\frac{r^2}{l^2}\right)-\left(1-\frac{M_0}{4\pi r}+\frac{r^2}{2l^2}\right)^{-1}\vec{v}^2\right]^{-1/2}
\end{equation}
Conservation of energy gives,
\begin{align}
&\left(1-\frac{M_0}{4\pi r}+\frac{r^2}{l^2}\right)\left[\left(1-\frac{M_0}{4\pi r}+\frac{r^2}{l^2}\right)-\left(1-\frac{M_0}{4\pi r}+\frac{r^2}{2l^2}\right)^{-1}\vec{v}^2\right]^{-1/2}=\notag\\&\left(1+\frac{r^2}{l^2}\right)\left[\left(1+\frac{r^2}{l^2}\right)-\left(1+\frac{r^2}{2l^2}\right)^{-1}\vec{\beta}^2\right]^{-1/2}\label{ae}.
\end{align}
Carrying out the expansion of (\ref{ae}) up to ${\cal{O}}(M)$ and ${\cal{O}}(1/l^2)$, and then comparing with the second equation of (\ref{asymrv}) we get,
\begin{equation}
C_{\pm}=\mp \frac{M_0(1-3\vec{\beta}_{\pm}^2)}{8\pi\vert \vec{\beta}_{\pm}\vert^3}.
\end{equation}
Thus $C_{\pm}$ do not get any contribution from the cosmological constant. Plugging $C_{\pm}$ in the soft factor contribution from the Schwarzschild part $\tilde S_{\rm gr}^M$ (\ref{Scpart}) we rightly reproduce the well known result of ``Classical Soft Theorem" in asymptotically flat spacetime,
\begin{align} \label{e}
&\tilde S_{\rm gr}\vert_{Sch} = -m\, \omega^{-1}\, \epsilon^{ij}\, 
\left\{ {1\over 1 - \hat n.\vec \beta_+} \, {1\over \sqrt{1-\vec \beta_+^2}}\, \beta_{+ i} 
\beta_{+ j} - {1\over 1 - \hat n.\vec \beta_-} \, {1\over \sqrt{1-\vec \beta_-^2}}\, \beta_{- i} 
\beta_{- j}\right\}\nonumber \\
&- i\, {m} \,   \ln \omega^{-1} \, \epsilon^{ij}\, \left[ 
{1\over \sqrt{1-\vec \beta_+^2}}\, \beta_{+ i} 
\beta_{+ j} \left\{C_+ {1\over 1-\hat n.\vec\beta_+}\right\}  - {1\over \sqrt{1-\vec \beta_-^2}}\, \beta_{- i} 
\beta_{- j} \left\{C_- {1\over 1-\hat n.\vec\beta_-} 
\right\}
\right]\, .
\end{align}

Equation (\ref{fr}) is the main result of this paper. It gives first order  corrections to Classical Soft theorem due to presence of a small cosmological constant. 

\section{Conclusions and Future Directions}\label{sc8}

In this paper we have studied the Classical Soft Theorem for asymptotically AdS spacetime in four spacetime dimensions. Our results can be trivially extended to higher spacetime dimensions. We computed the radiation profile produced in a classical scattering process in AdS Schwarzschild background in a probe scattering approximation. Our analysis assumes the cosmological constant $\Lambda= -\frac{3}{l^2}$ as a perturbation parameter over asymptotically flat gravity and all results are valid to leading order in $\frac{1}{l^2}$. The reason for treating $\Lambda$ perturbatively in our analysis is the following : in the classical scattering computations that we have performed, one prime consideration is that the scattering takes place within a neighbourhood of finite radius (say $a$) from the chosen origin. The detector, that traces the gravitational wavefront produced in the scattering process, is placed at a far away point $\vec x : R= |\vec x|$ and we work in large R limit. The system can go through any possible interactions within the region of radius $a$ and out side this region it is only gravitational interaction that plays the dominant role. In an asymptotically AdS spacetime, this assumption does not hold true in general. Since $AdS_4$ comes with a spatial boundary, in order to solve the Einstein's equations in this background, we must also specify the boundary conditions for the associated fields. The most used boundary condition that is imposed is a reflective boundary condition. As a result the reflected waves bounce back from the boundary, they interact again non-trivially and they contribute to net perturbation. Thus, we would not be able to define a region of radius $a$ to contain all possible interactions and make the system interact only gravitationally out side of it. 

 To avoid this issue we have studied $AdS_4$ in isotropic coordinates (\ref{mm}) and thus it comes with an unique problem. Since, the  coordinate system, only perturbatively (in $\Lambda$) covers the spacetime, the spatial boundary of $AdS_4$ is not a part of this co-ordinate system. Thus we can, in principle, demand that the perturbations die sufficiently fast at large spatial distances for this co-ordinate system ($e_{ij} \rightarrow 0$ as $r,l \rightarrow \infty : \frac{r^2}{l^2}<<1$). We have assumed the boundary condition similar to \cite{PhysRevD.18.3565}, where causal AdS spacetime was thought of as embedded in the Einstein static universe. The  non-trivial boundary condition at spatial boundary of our underlying $AdS_4$ spacetime are evident in equations (\ref{Gl}) and (\ref{e26}-\ref{e2end}). The profile does not involve Theta functions, as in usual with reflective boundary conditions. This boundary condition ensures that in our calculations we only need to take into account the bulk-to-bulk Green's function
  to compute the radiation profile at the point of interest. We do not consider the contributions from waves reflected from the boundary to it. In terms of usual global coordinates, the long range force in AdS produces an extra effect (as compared to the flat case) to the particle trajectory as given in (\ref{tlr}), but in perturbative computation it does not play any role. In other words, the reflective boundary conditions become important to higher order in perturbative computations. It will be nice to perform a classical radiation computation to the next order in $\Lambda$ or possibly a non-perturbative computation to capture this effect, but it looks technically difficult at this point.
  
 We end this paper with some interesting open questions:
 \begin{itemize}
 \item 
In \cite{Compere:2019bua} Compere et.al. have presented a boundary condition for $AdS_4$, which gives rise to a non-trivial asymptotic symmetry group at the boundary, namely $\Lambda-BMS_4$. It would be interesting to see, whether our boundary condition on fields is consistent with theirs and if even in $AdS_4$, a relation between boundary symmetry group and soft radiation can be found.
\item 
In \cite{He:2014laa}, it was argued that, for asymptotically flat spaces, the observational consequences of BMS symmetry are embodied in the soft factorization of graviton scattering amplitudes as Weinberg soft graviton theorem is essentially a rewriting of the formula for gravitational memory \cite{Strominger:2014pwa}. Memory effect is characterised by the difference between the proper displacement between the observers before and after a  gravitational waves passes by their locations. As was discussed in the original paper by Strominger et. al. \cite{Strominger:2014pwa}, this relation requires a particular class of gravitational waves that Braginsky and Thorne refers to as ``Bursts with Memory". A simple particle scattering calculation may not produce these effects at all. In fact, in a recent paper \cite{Chu:2019ssw} such a calculation was done in AdS background and no observable memory effect was found after the passage of gravitational waves. Only the region with a non-zero gravitational field showed to have a observable displacement of geodesics. On the contrary, since we have studied radiation in a AdS space where the cosmological constant is small, it is possible connect our work with the flat space results of Strominger et.al. This computation is equally applicable for our universe which has a small dS potential and hence may turn out to be extremely important from observational perspective. We hope to report on this in near future.

\item 
In \cite{Saha:2019tub}, the authors derived the classical soft photon theorem from first principle. They further generalised the soft graviton theorem by allowing electromagnetic interactions among the incoming and the outgoing particles. It will be nice to see how these results get modified in presence of a small cosmological constant in the background.

\end{itemize}

\vspace{1cm}
{\bf Acknowledgements}\\

We would like to thank Sayali Bhatkar, Sayantani Bhattacharyya, Karan Fernandes, Alok Laddha, Debangshu Mukherjee, Arnab Rudra, Biswajit Sahoo and Ashoke Sen for useful discussions. AM would like to thanks HRI for hospitality during the initial course of this work. AB would like to thank IISER Bhopal for their hospitality during the entirety of this project. Our work is partially supported by a SERB ECR grant, GOVT of India. Finally, we thank the people of India for their generous support to the basic sciences.

\appendix
\section{Connection, Riemann, Ricci terms}\label{AppA}
The non-vanishing connection terms are
\begin{align}
&\Gamma^0_{0i}=\phi_i+\frac{1}{2l^2}\rho^2_i;\quad \Gamma^i_{00}=\phi_i+\frac{1}{2l^2}\rho^2_i\notag\\&\Gamma^i_{jk}=\left(-\delta_{ij}\phi_k-\delta_{ik}\phi_j+\delta_{jk}\phi_i\right)+\frac{1}{4l^2}\left(\delta_{ij}\rho^2_k+\delta_{ik}\rho^2_j-\delta_{jk}\rho^2_i\right)
\end{align}
where $\phi_i=\partial_i\phi$ and $\rho^2_i=\partial_i\rho^2$ and terms are retained till ${\cal{O}}(1/l^2)$.

The Riemann, Ricci tensor and Ricci scalar are as follows
\begin{align}
R^0{}_{ij0}&=\phi_{ij}+\frac{1}{2l^2}\rho^2_{ij}\notag\\R^k{}_{ijm}&=\left(-\delta_{km}\phi_{ij}+\delta_{im}\phi_{kj}+\delta_{kj}\phi_{im}-\delta_{ij}\phi_{km}\right)+\frac{1}{4l^2}\left(\delta_{km}\rho^2_{ij}-\delta_{im}\rho^2_{kj}-\delta_{kj}\rho^2_{im}+\delta_{ij}\rho^2_{km}\right)\notag\\R_{00}&=\phi_{ii}+\frac{1}{2l^2}\rho^2_{ii}\qquad R_{ij}=\delta_{ij}\phi_{kk}-\frac{1}{4l^2}\left(\delta_{ij}\rho^2_{kk}+3\rho^2_{ij}\right)\notag\\R&=2\phi_{ii}-\frac{2}{l^2}\rho^2_{ii}
\end{align}
\section{Analysis of soft expansion for $\tilde{e}^{(5)}_{ij}$, $\tilde{e}^{(6)}_{ij}$ and $\tilde{e}^{(7)}_{ij}$}\label{AppB}
To study the soft expansion of $\tilde{e}^{(5)}_{ij}$ we have to first evaluate the action of $\nabla_i=\frac{\p}{\p x^i}+\frac{\p}{\p r'^i}$ on $G_l$ of \ref{gl}. Using
\begin{align}
\p_i G_l=\left(i\omega \hat{n}_i+\frac{\hat{n}_i}{R}\right)G_l,\quad \p_i'G_l=i\omega \hat{n}_i G_l-\frac{4i}{\omega}e^{i\omega(R-\hat{n}.\vec{r}')}R \hat{n}_i
\end{align}
we can write 
\begin{align}
\nabla_i G_l=-\frac{4i}{\omega}&e^{i\omega(R-\hat{n}.\vec{r}~')}R \hat{n}_i\notag\\\nabla_i\nabla_j G_l=-\frac{8i}{\omega}&e^{i\omega(R-\hat{n}.\vec{r}~')}\hat{n}_i\hat{n}_j,\,\nabla_k^2 G_l=-\frac{8i}{\omega}e^{i\omega(R-\hat{n}.\vec{r}~')}
\end{align}
 Most of the term from this part of $\tilde{e}^{(5)}_{ij}$ vanishes while contracting with polarization tensor $\epsilon^{ij}$ since $\epsilon^{ij}\hat n_j=\epsilon^{ij} k_j/|\vec k|=0$. The only term we need to look at carefully is
\begin{align}
T_5=& -\frac{m}{8\pi l^2}\int dt \frac{dt}{ds}e^{i\omega t}v_iv_j\nabla_k\nabla_k G_l\bigg|_{\vec{r}~'=\vec{r}(t)}, \nonumber\\
=& i\frac{m}{\pi l^2 \omega}\int dt \frac{dt}{ds}e^{i\omega t}v_iv_j~e^{i\omega (R - \hat{n}.\vec{r}~)} 
\end{align}
Following \ref{eas} and doing an integration by parts one can get,
\begin{align}
T_5=& -\frac{m}{\pi l^2 \omega^2}\int dt \frac{d}{dt}\left[\frac{1}{\sqrt{1-\vec{v}(t)^2}}\frac{1}{1- \hat{n}.\vec{v}}v_iv_j\right]e^{i\omega (R+t - \hat{n}.\vec{r})}\label{t5}
\end{align}
Putting the expressions of $v$ s one can simplify \ref{t5} as,
\begin{align}
=& -\frac{m}{\pi l^2 \omega^2}\int dt \frac{1}{\sqrt{1-\vec{\beta}_{\pm}(t)^2}}\frac{1}{1- \hat{n}.\vec{\beta}_{\pm}}\beta_i\beta_j \frac{d}{dt}\left[1-\#1/ t\right]e^{i\omega (R+t - \hat{n}.\vec{r})}
\label{66}
\end{align}
The integration in \ref{66} using \ref{i1} will give $\omega$ independent term and $O(\omega\ln{\omega}^{-1})$ terms with the $\frac{1}{\omega^2 l^2}$ factor in the front. Hence they are finite in $\omega\rightarrow 0$ limit keeping $\omega l$ fixed and therefore do not contribute to the soft factor.

Next we will look into the soft expansion of $\tilde{e}^{(6)}_{ij}$. It do not involve any derivative on $G_l$. 
\begin{align} 
\tilde{e}^{(6)}_{ij}(\omega, \vec{x}) =& -\frac{m}{2\pi \, l^2} R e^{i\omega R} \int dt e^{i\omega (t-\hat{n}.\vec{r}(t))} \frac{d}{dt}\left[\frac{dt}{ds} v_{i}v_{j}\hat{n}.\vec{r}'(t)\frac{1}{1-\hat{n}.\vec{v}(t) } \right]\\
\\=&- \frac{m}{2\pi \, l^2} R e^{i\omega R} \int dt e^{i\omega (t-\hat{n}.\vec{r}(t))} \frac{d}{dt}\left[ \,a_1 t - a_2 \ln t \right] \frac{\beta^i_{\pm}\beta^j_{\pm}}{1-\hat{n}. \vec{\beta}_{\pm}}(\hat{n}. \vec{\beta}_{\pm})\frac{1}{\sqrt{1-\beta_{\pm}^2}}
\end{align}
where,
\begin{align}
a_1 =& 1\\
a_2 =& -c_{\pm}\beta_{\pm}
\end{align}
Doing the following manipulation
\begin{align} 
&\tilde{e}^{(6)}_{ij}(\omega, \vec{x}) = \frac{m}{2\pi \, l^2} R e^{i\omega R} \int dt e^{i\omega (t-\hat{n}.\vec{r}(t))} \frac{d}{dt}[a_1 \, t + a_2 \ln t^{-1} ] \frac{\beta^i_{\pm}\beta^j_{\pm}}{1-\hat{n}. \vec{\beta}_{\pm}}(\hat{n}. \vec{\beta}_{\pm})\frac{1}{\sqrt{1-\beta_{\pm}^2}}\\
=& -\frac{m}{2\pi \, l^2} R e^{i\omega R} \int dt \frac{1}{i\omega}e^{i\omega (t-\hat{n}.\vec{r}(t))} \frac{d}{dt}[a_1 \, +\frac{1}{t} (-a_2- a_1 c\, \alpha)  ] \frac{\beta^i_{\pm}\beta^j_{\pm}}{(1-\hat{n}. \vec{\beta}_{\pm})^2}(\hat{n}. \vec{\beta}_{\pm})\notag\\
=&  \frac{im}{2\pi \, l^2} R e^{i\omega R} \frac{\beta^i_{\pm}\beta^j_{\pm}}{(1-\hat{n}. \vec{\beta}_{\pm})^2}\frac{\hat{n}. \vec{\beta}_{\pm}}{\sqrt{1-\beta_{\pm}^2}}\notag\\&\left\lbrace \frac{1}{\omega} (a_{1+}- a_{1-})+ i \ln \omega^{-1} \{a_{1-} (a_{2-}+c\, a_{1-}\,  \alpha_{-}) - a_{1+} (a_{2+}+c\, a_{1+}\,  \alpha_{+}) \} \right\rbrace\notag\\
=&  \frac{im}{2\pi \, \omega^2 l^2} R\frac{\beta^i_{\pm}\beta^j_{\pm}}{(1-\hat{n}. \vec{\beta}_{\pm})^2}\frac{\hat{n}. \vec{\beta}_{\pm}}{\sqrt{1-\beta_{\pm}^2}} e^{i\omega R}\left\lbrace \omega (a_{1+}- a_{1-})+ i \omega^2\ln \omega^{-1} \right.\notag\\&\left.\left\{a_{1-} (a_{2-}+c\, a_{1-}\,  \alpha_{-}) - a_{1+} (a_{2+}+c\, a_{1+}\,  \alpha_{+}) \right\} \right\rbrace \label{e66}
\end{align}
we can see that in \ref{e66} all terms are finite in $\omega\rightarrow 0$ limit.

Doing the same analysis for $\tilde e^{(7)}_{ij}$ we get 
\begin{align}
\tilde e^{(7)}_{ij}(\omega, \vec{x}) &=-i\omega\frac{5m}{64\pi l^2}\int dt \frac{dt}{ds} e^{i\omega t}(v_i \nabla_j+v_j\nabla_i)G_l\notag\\&=\frac{5m}{16\pi l^2}\int dt \frac{dt}{ds} e^{i\omega(R+ t-\hat{n}.\vec{r}})(v_i \hat{n}_j+v_j \hat{n}_i)\label{77}
\end{align}
The contribution to the integral from \ref{77} vanishes after $\tilde e^{(7)}_{ij}$ is contracted with the polarization tensor $\epsilon^{ij}$, as $\epsilon^{ij}\hat n_j=\epsilon^{ij} k_j/|\vec k|=0$. 

\end{document}